\documentclass[useAMS,usenatbib,usegraphicx]{mn2e}
\def\la{\raise.5ex\hbox{$<$}\kern-.8em\lower 1mm\hbox{$\sim$}}
\def\ma{\raise.5ex\hbox{$>$}\kern-.8em\lower 1mm\hbox{$\sim$}}

\def\msol{M$_{\odot}$ }

\def\kms{$\rm km\, s^{-1}$}
\def\cm3{$\rm cm^{-3}$}
\def\Ts{$\rm T_{*}$~}
\def\Vs{$\rm V_{s}$~}
\def\n0{$\rm n_{0}$}
\def\B0{$\rm B_{0}$}

\def\erg{$\rm erg\, cm^{-2}\, s^{-1}$}
\def\mum{$\mu$m~}

\def\L12{L$_{12\mu m}$~}
\def\F12{F$_{12\mu m}$~}

\def\Hb{H$\beta$~}
\def\Ha{H$\alpha$~}
\def\Ly{Ly$\alpha$~}

\title[Spectra from the turbulent GC]{Spectra from the shocked  nebulae revealing turbulence near the  Galactic Centre }
\author[M. Contini and I. Goldman]{ M. Contini$^{ 1}$ and I. Goldman$^{2 }$\thanks{Itzhak Goldman thanks  the Department of Astronomy and Astrophysics, Tel Aviv University, for
the hospitality while on Sabbatical Leave from Afeka College.}   
\\
$^{1 }$School of Physics and Astronomy, Tel Aviv University, Tel Aviv
69978, Israel \\
$^{2 }$Afeka,
Tel Aviv Academic College of Engineering,
  Tel Aviv 69107, Israel, email:
goldman@afeka.ac.il \\
}

\begin{document}

\date{Accepted: Received ; in original form 2010 month day}

\pagerange{\pageref{firstpage}--\pageref{lastpage}} \pubyear{2009}

\maketitle

\label{firstpage}

\begin{abstract}

The  spectra emitted from clouds near the Galactic Centre (GC) are investigated
calculating the  UV-optical-IR lines   using the physical  parameters and the  
element abundances obtained by the detailed modelling of  mid-IR  observations. 
The graphical presentation  of the spectra   reveals the  strong 
lines. 
The characteristic  line ratios calculated at the nebula  provide
information about  the dereddened spectra emitted from regions near the GC.
These line ratios 
 are compared with those observed in active galaxies.
We have found  that the  physical conditions in the nebulae near the GC are different
from those of starburst galaxies and  AGN, namely, gas velocities and densities as well as the
photoionization  fluxes are relatively low. 
 The geometrical thickness of the emitting filaments is particularly small suggesting 
that  matter is strongly fragmented
by instabilities leading to an underlying shock-generated  turbulence.
This  is revealed by the power spectra of the radial velocities, of the  mid-IR continuum
flux and of the computed Si/H relative abundances. Moreover,  turbulence  could  amplify the
initial magnetic fields.

\end{abstract}


\section{Introduction}

The  Galaxy   should be the largest information source about high and low-ionization
level lines,  however the  central regions  cannot be observed  in the optical and UV range
 because of strong extinction (Schultheis et al. 2009). 

 Spectroscopic data  in the infrared, UV,  and in the X-ray range were lacking`
before the Spitzer, FUSE, and XMM era, respectively.
In  recent years,
 Spitzer infrared  spectra observed by Simpson et al (2007) and references therein  and modelled in detail  by
 Contini (2009, hereafter Paper I)
 allowed  a detailed investigation of the gas and dust
structures  near the Galactic Centre (GC).
The  most serious problem  was to  obtain the  spectra that had been properly
corrected for extinction.

Fig. 1 (top) shows a radio image of this region (Yusef-Zadeh \& Morris 1987). 
A sketch of the entire Galaxy (Nakanishi \& Sofue 2006) is  given in Fig. 1 (bottom)  
 in order to   emphasize the relatively small  extent of the observed region.
The  Sgr A West HII region  contains a quiescent black hole $\sim$ 4 10$^{6}$ \msol (Ghez et al.
 2005; Eisenhauer et al. 2005) which is coincident with the radio source Sgr A* and is located
at the GC. It also contains  a cluster of massive stars.
Two  other  clusters of  young massive stars and massive molecular clouds
(Sch\"{o}del et al. 2006) appear in the GC,  the Arches Cluster and the Quintuplet Cluster
located $\sim$  25 pc away in the plane of the sky.
The very massive Arches Cluster (Nagata et al 1995 and Cotera et al. 1996)  
 of young stars  heats and ionizes the  region of the
Arched Filaments  and the Quintuplet Cluster ionizes the Sickle.

The star flux from the Quintuplet Cluster  affects the clouds in the extended region  including  the Bubble.
A detailed description of the GC is given by Simpson et al (2007).

\begin{figure}
\begin{center}
\includegraphics[width=0.40\textwidth]{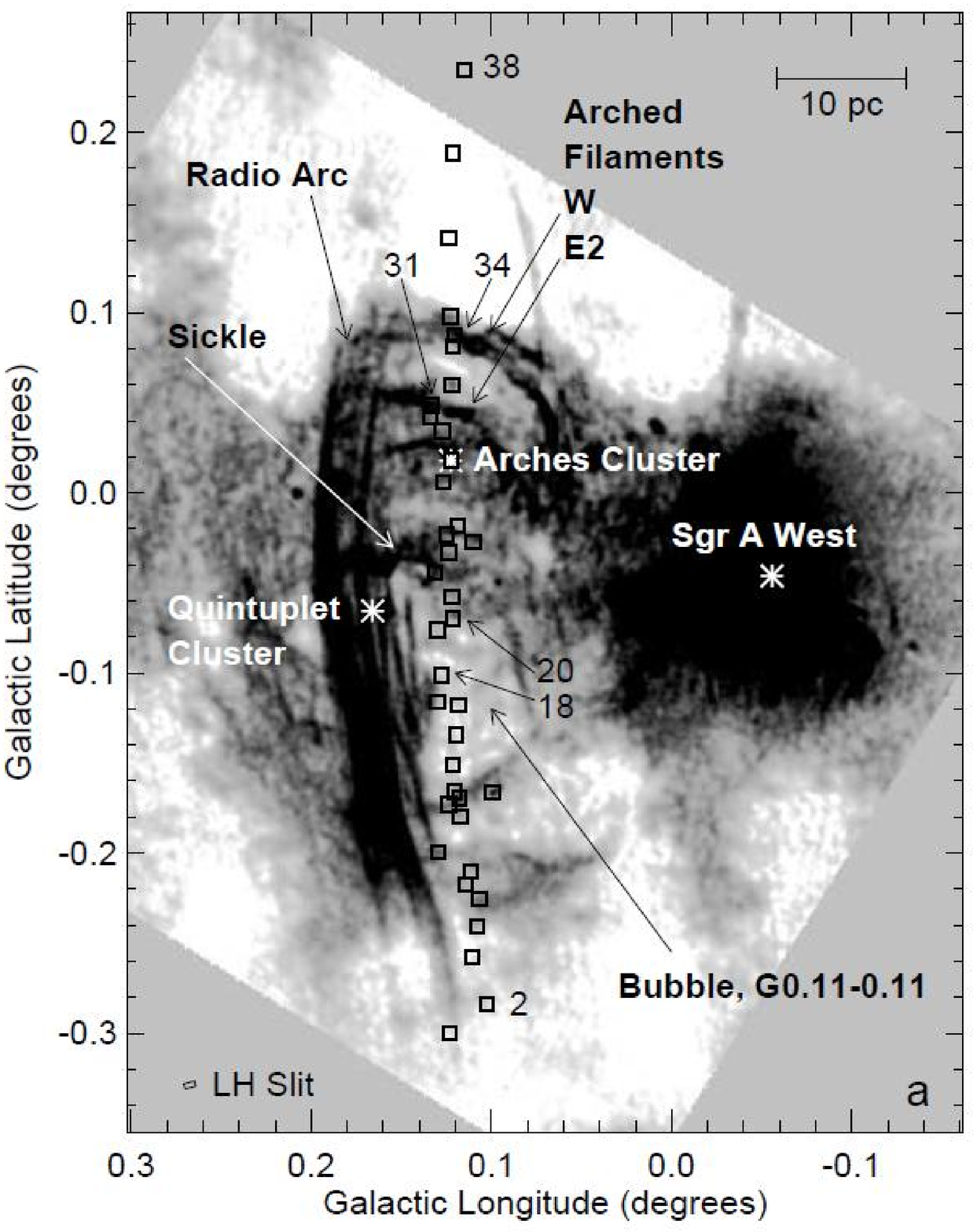}
\includegraphics[width=0.45\textwidth]{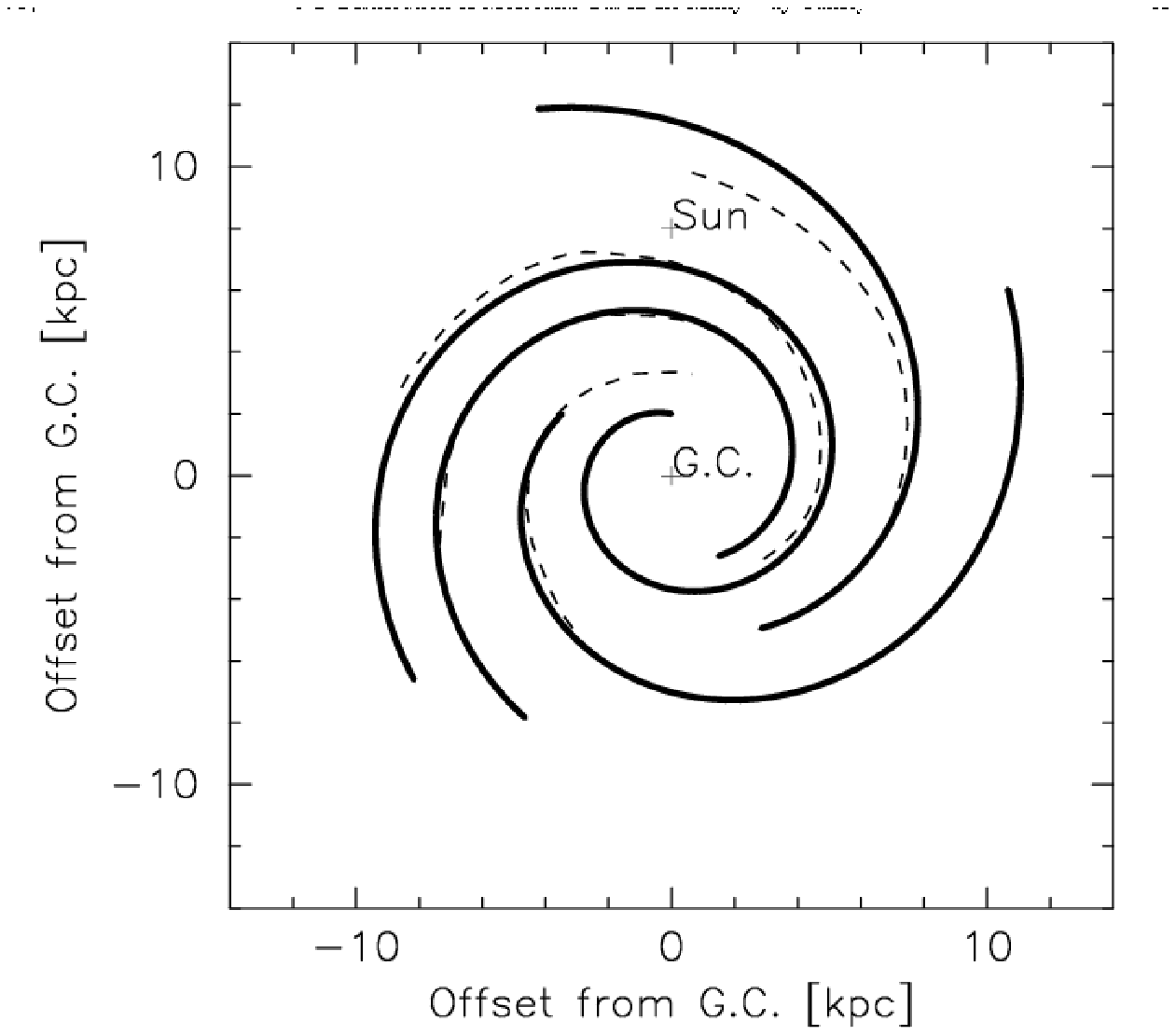}
\caption{Top : Image of the Galactic Center with the observed positions as square boxes 
upon the radio continuum imaged at 21 cm by Yusef-Zadeh \& Morris (1987)
adapted from Simpson et al (2007, fig. 1).
Bottom :  The sketch of the Galaxy spiral arms is adapted from Nakanishi \& Sofue (2006), fig. 15}
\end{center}
\end{figure}

The spectra were observed by Simpson et al.
 with the Spitzer IRS instrument, at 38 locations scanning across the Galactic plane; 
the locations and largest observed aperture size are presented in Fig. 1 (top).   Accurate
modelling  and the precision of the data fit (Paper I)
allowed   evaluation of the physical
conditions of the emitting gas, such as
 temperatures, densities, velocities,  the magnetic field and the relative abundances of the elements.
Moreover, the spectra contain  a number of lines from different excitation levels
and from different elements enough to constrain the models.
This  suggests that the same  models  which could  reproduce  the observational data in the IR
could reproduce  the data also in other ranges, e.g. optical, UV, etc.

So we  present in Tables 1-4  model results for lines
which were not included in Paper I. 
Most of the lines, particularly in the UV,   cannot be observed because 
of extinction.
Our aim is to provide a tool for correction once the spectra will be
observed at least as upper limits. 

We adopt  a simple graphical  method
which  shows   predicted dereddened spectra and permits a rapid  comparison  with 
the spectra observed in other galaxies.
The most  significant lines
throughout  a large spectral range,  from SiIII 1206 close to the \Ly line up to the FIR,
 [CII] 156 and [NII] 203, are selected.  The UV-optical lines are  in \AA,
while the IR lines are  in \mum throughout all the paper.
The  spectra are presented in Sect. 2.

We concluded in Paper I that the spectra observed in the nebulae near the GC     
show the characteristics of
HII regions. However,   an  active nucleus, even if very weak is  not excluded.
So we will  compare the mid-IR   line ratios emitted from nebulae near  the GC with those of LINERs,
AGN, starburst galaxies, etc.
 
Some significant  line ratios  observed in the IR  are analysed  in Sect. 2.1 and 2.2.
In the optical and UV ranges we will use model results which are constrained by the mid-IR observations. 
In Sects. 2.3  and 2.4 the line ratios calculated  for nebulae near the GC   are   
compared with the set of  spectra observed  at different positions throughout   single  galaxies  
and with the ensemble of  line ratios observed from a sample of active galaxies (starburts, AGN, etc.).
The  model calculations account for photoionization and for shocks. In Sect. 3 the results are
discussed in the light of  shock generated turbulence.
Concluding remarks follow   in Sect. 4.

\section{The line spectra}

The spectra  are calculated adopting the code
SUMA (see http://wise-obs.tau.ac.il/$\sim$marcel/suma/index.htm for
a detailed description),
 that simulates the physical conditions of an
emitting gaseous cloud
under the coupled effect of photoionisation from an external radiation
source and shocks. Both line and continuum emission from the gas
are  calculated consistently with dust reprocessed radiation,
 in a plane-parallel geometry.

The input parameters are: the  shock velocity \Vs, the   preshock density \n0,
the preshock magnetic field \B0, the colour  temperature of the hot star \Ts,
the ionization parameter $U$ for black body (BB) dominated fluxes,  or the power-law (pl)
flux  reaching the nebulae
from the active center $F_h$  in number of photons cm$^{-2}$ s$^{-1}$ eV$^{-1}$ at the Lyman limit.
The geometrical thickness of the emitting nebula $D$,
the dust-to-gas ratio $d/g$, and the  abundances of He, C, N, O, Ne, Mg, Si, S, A, Fe
relative to H are also accounted for.
Solar abundances are adopted (Allen 1976) in the models presented by  Contini \& Viegas (2001a,b).
The distribution of the grain radius downstream
is determined by sputtering,  beginning with an initial  radius of 0.2 \mum.

We refer to the regions observed in 38 positions near the GC by Simpson et al. (2007).
The  absolute intensities of the most important lines (Tables 1-4) emerging from
a zero intensity continuum are shown in  graphical form in Fig. 2.
The spectra   show  the  line intensities  but they
do not include the  line  profiles.

 This representation is not aimed to  measure or  compare in detail the line intensities.
That  should  be done using  the Tables 1-4. We are interested in
showing the trend, e.g. the   rising and fading of some important lines close to the
clusters and in the ISM.
 The spectra from the positions  close to 17-25 show the photoionizing  effect from the Quintuplet Cluster,
while the filaments from positions  close to 27-35 are photoionized by the Arches Cluster.

Notice  that [NII] 203 appears in some intermediate
postitions (31-35, 10-13, 23-26), while [CII] 156 is always present, explaining  its  detection
in many other galaxies.
Moreover, the UV and mid-IR lines are all of comparable strength. The spectra are very different from
those of starburst galaxies (Fig. 4) calculated with higher \Vs, and \n0  which are dominated by lines
 in the optical range.
The spectra corresponding to LINER conditions are strong in the same ranges as those of
the GC, but the line ratios are very different.

The graphical presentation   allows for easy recognitions of the  lines strong
enough to be observed  in the GC (IR spectra) and
 in other galaxies (UV-optical-IR spctra) and therefore useful for comparison purposes.

 The corresponding models indicate the characteristic conditions
of the emitting gas.
The models adopted for the calculations which lead to the best fit of the mid-IR observations
are given  in Paper I (table 2 and fig 2).

%\flushleft
\begin{figure*}
\includegraphics[width=0.49\textwidth]{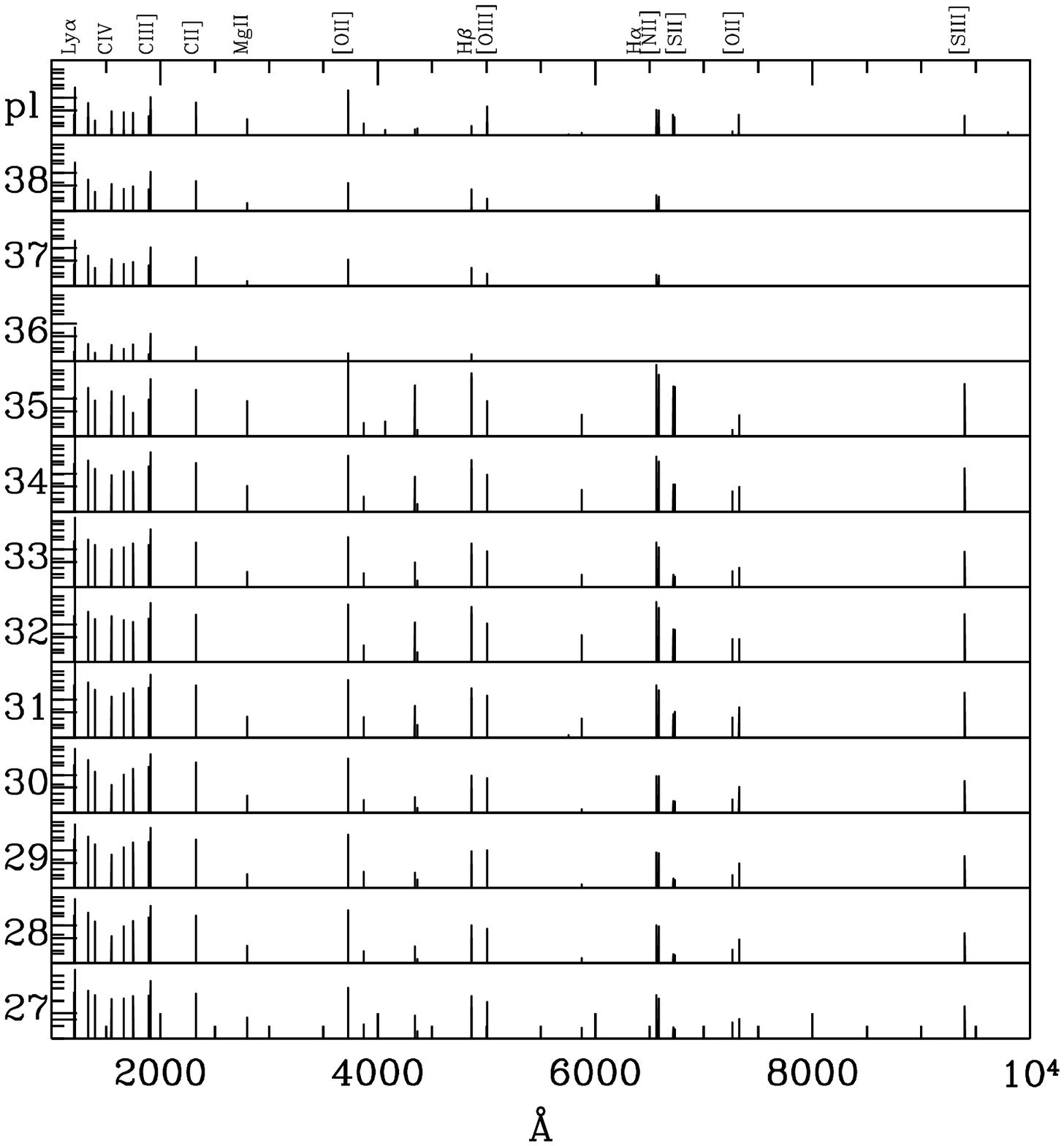}
\includegraphics[width=0.49\textwidth]{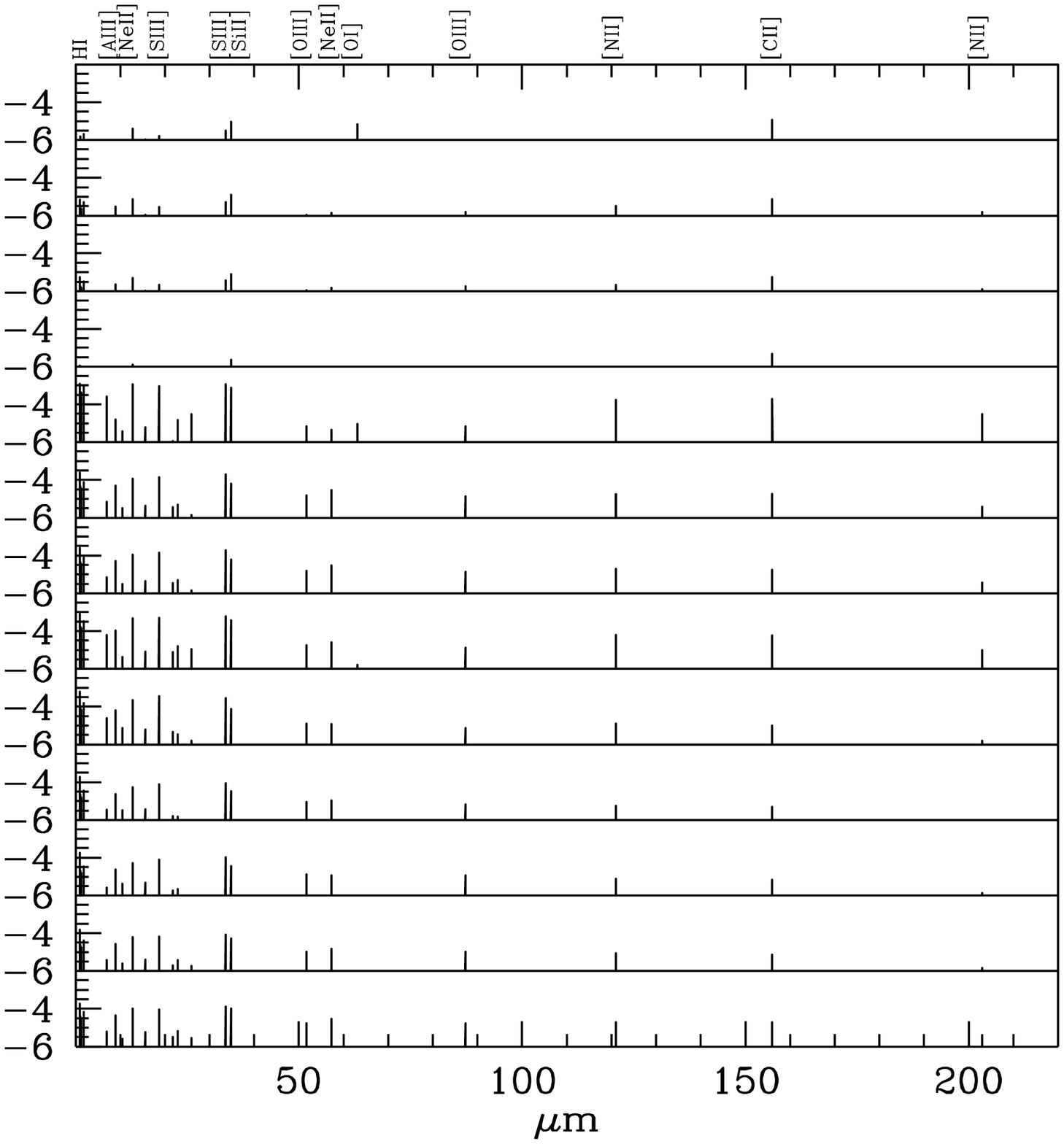}     
\includegraphics[width=0.49\textwidth]{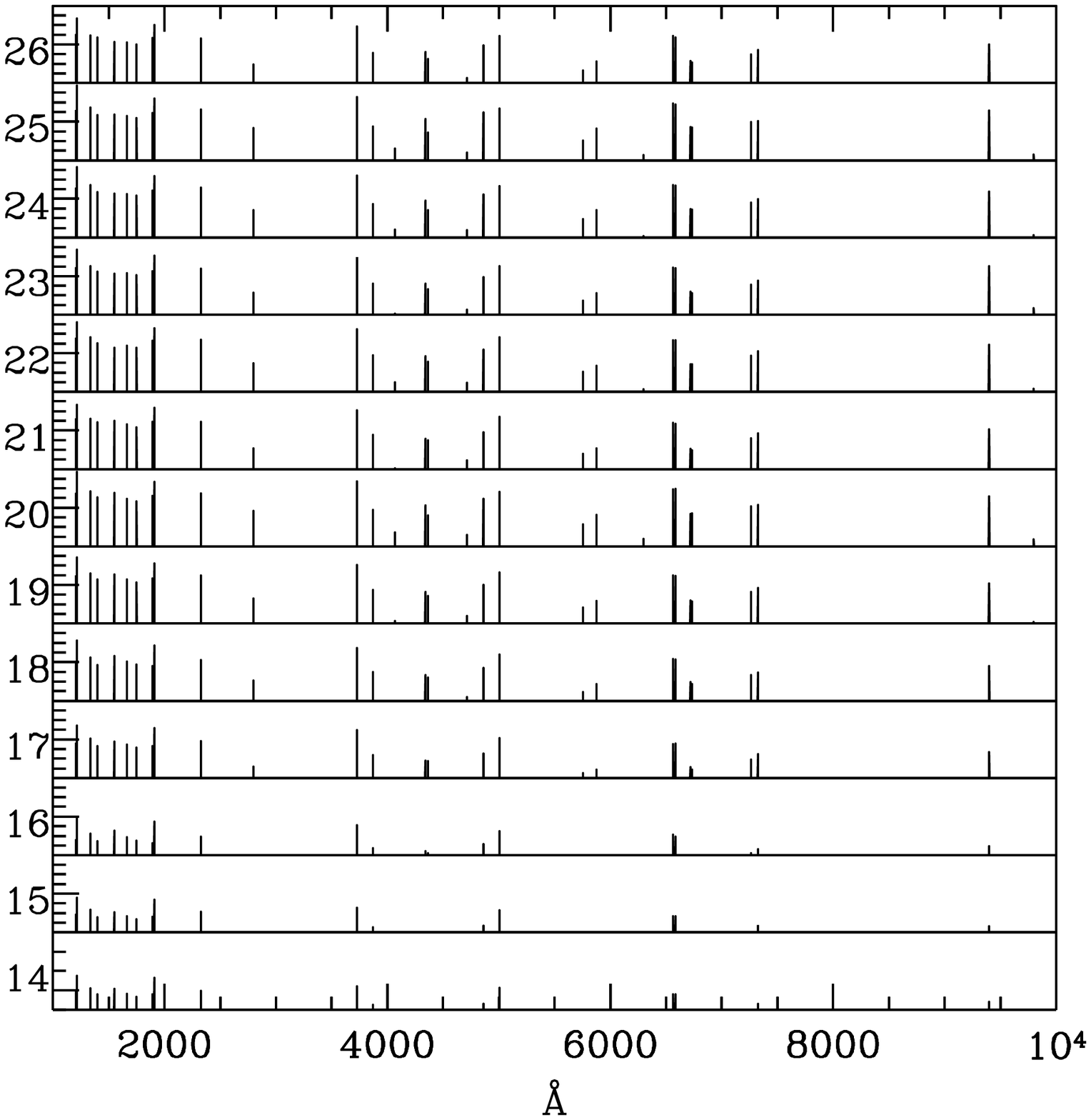}
\includegraphics[width=0.49\textwidth]{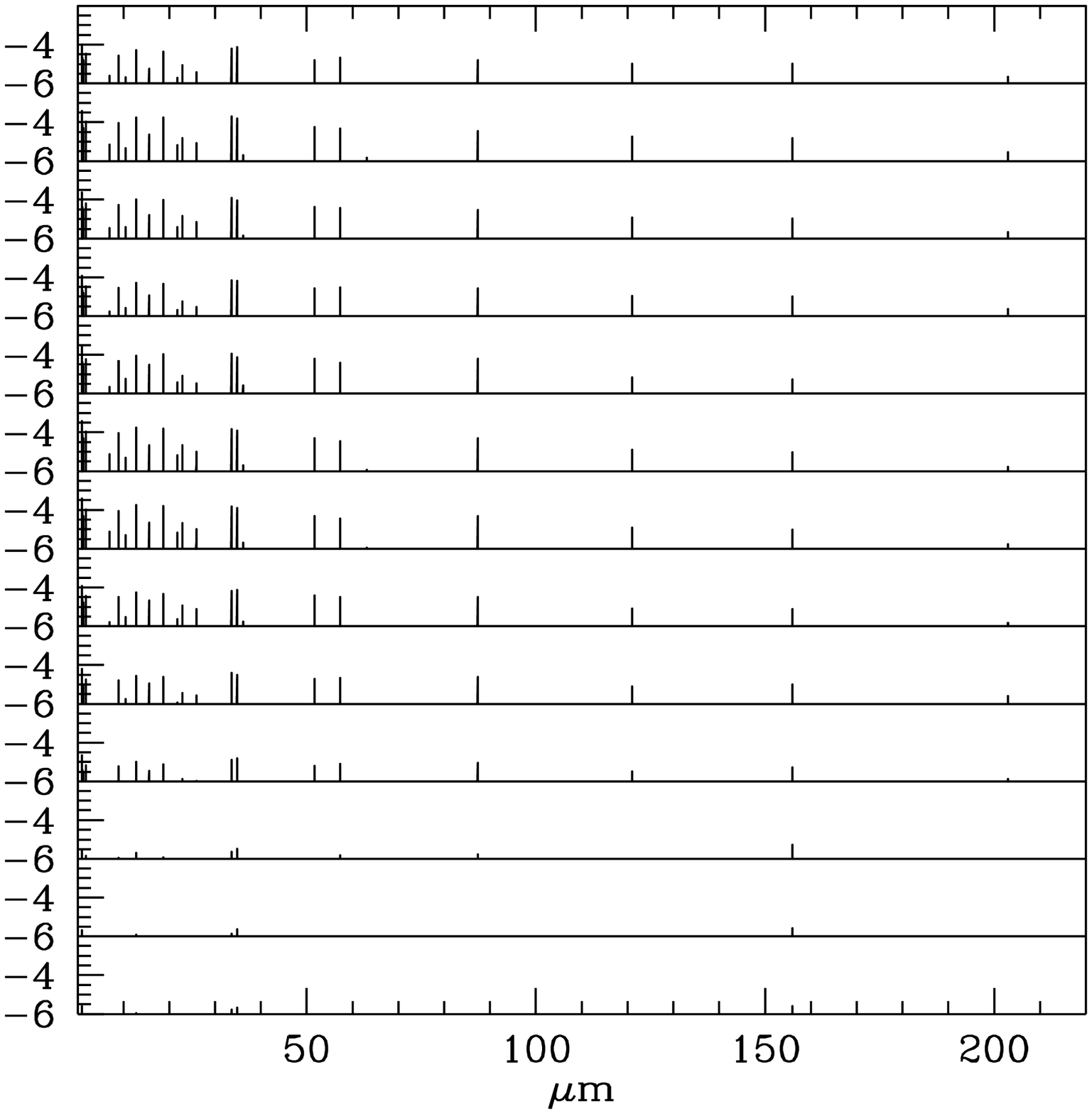}
\caption{UV-optical-IR lines calculated (in \erg)
by models m1-m38 and  m$_{pl}$ (Contini 2009) corresponding to the
observed positions (Simpson et al (2007).
}

\end{figure*}

\begin{figure*}
\includegraphics[width=0.49\textwidth]{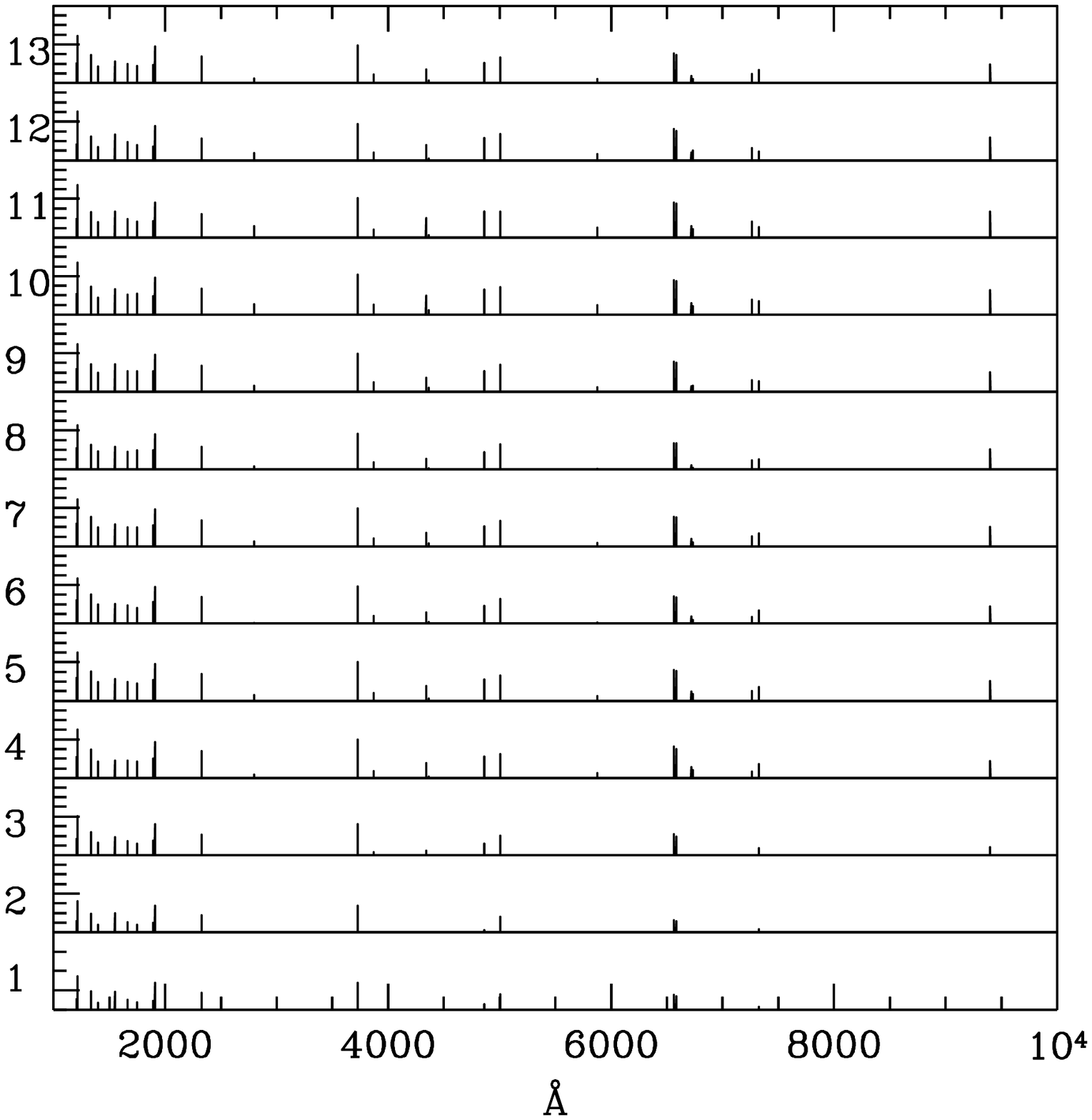}
\includegraphics[width=0.49\textwidth]{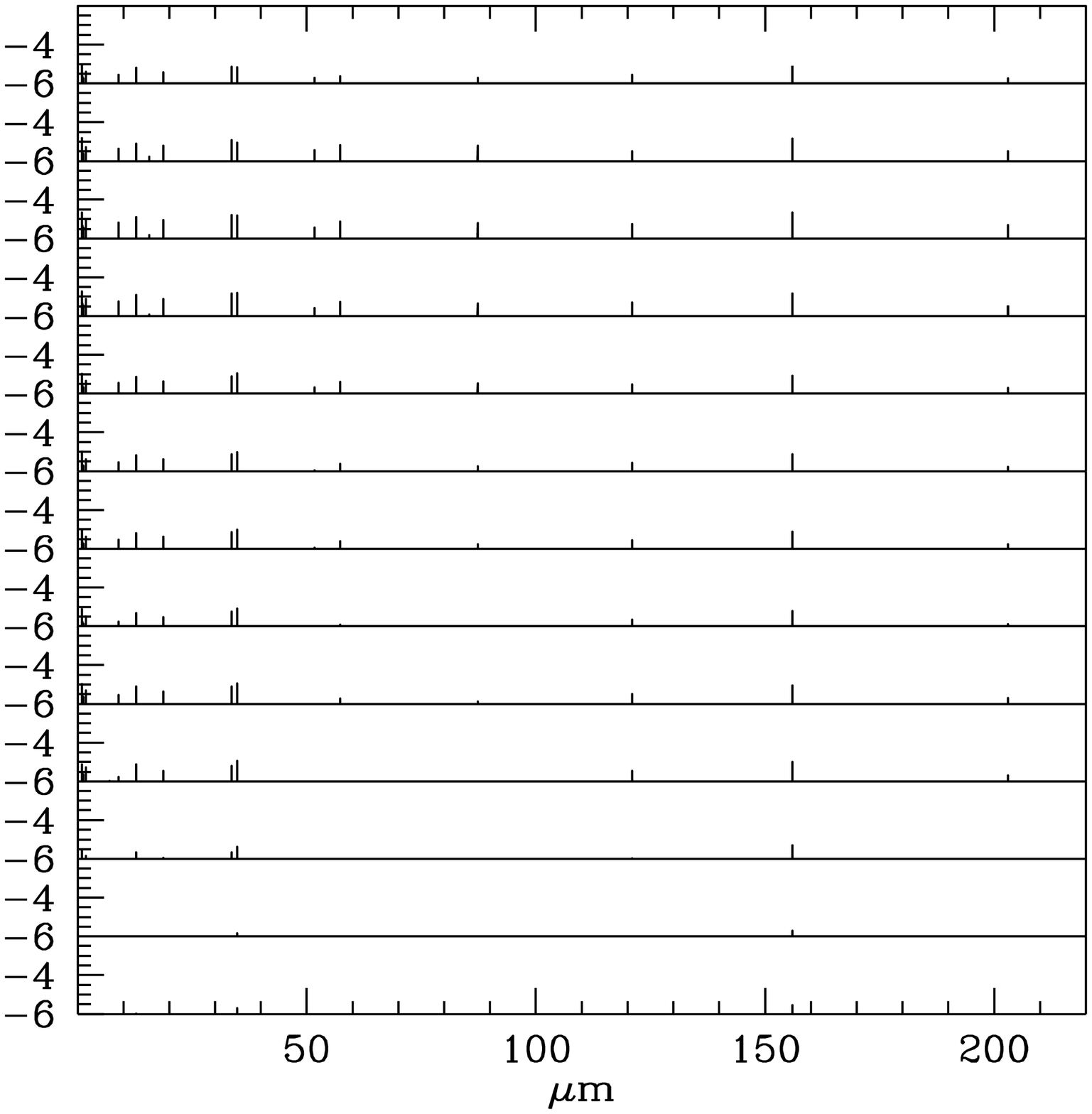}
\smallskip
\centerline{{\bf Figure 2} -- {\it continued}
}

\end{figure*}

\begin{table*}
\caption{UV lines (\erg)}
\tiny{
\begin{tabular}{ l l l l l l l l l l l ll} \\ \hline \hline
\ position & SiIII & \Ly & CII & SiIV& CIV & OIII] & NIII] & SiIII] & CIII] & CII] & MgII\\
\ $\lambda$ (\AA) &1206 & 1215 & 1335 & 1397 & 1549 & 1663 & 1750 & 1892 & 1909 & 2326& 2798 \\ \hline 
\     1&3.87D-6& 5.32D-5& 8.98D-6& 2.24D-6& 8.41D-6& 3.29D-6& 2.43D-6& 2.90D-6& 2.39D-5& 7.67D-6&2.9D-7\\
\     2&4.24D-6& 4.03D-5& 9.23D-6& 2.51D-6& 9.91D-6& 3.46D-6& 2.52D-6& 3.20D-6& 2.48D-5& 7.94D-6&4.7D-7\\
\     3&7.64D-6& 1.03D-4& 1.56D-5& 4.60D-6& 8.81D-6& 5.27D-6& 3.90D-6& 5.58D-6& 4.10D-5& 1.19D-5&4.3D-7\\
\     4&1.36D-5& 3.25D-4& 3.04D-5& 7.28D-6& 8.19D-6& 7.80D-6& 6.89D-6& 1.00D-5& 7.28D-5& 2.54D-5&1.5D-6\\
\     5&1.69D-5& 3.24D-4& 3.38D-5& 9.49D-6& 1.36D-5& 9.49D-6& 7.80D-6& 1.23D-5& 8.06D-5& 2.46D-5&2.0D-6\\
\     6&1.79D-5& 2.13D-4& 3.24D-5 &9.79D-6& 1.05D-5& 8.88D-6& 6.64D-6& 1.29D-5& 7.72D-5& 2.38D-5&1.1D-6\\
\     7&1.67D-5& 2.75D-4& 3.30D-5& 9.46D-6& 1.42D-5& 9.79D-6& 9.90D-6& 1.19D-5& 8.08D-5& 2.28D-5&1.8D-6\\
\     8&1.39D-5& 1.86D-4& 1.87D-5& 8.25D-6& 1.40D-5& 8.10D-6 &9.75D-6& 9.52D-6& 6.19D-5& 1.44D-5&1.4D-6\\
\     9&1.66D-5& 3.00D-4& 2.76D-5& 1.01D-5& 2.69D-5& 1.19D-5& 1.20D-5& 1.19D-5& 8.68D-5& 2.21D-5&2.1D-6\\
\    10&1.30D-5& 4.86D-4& 2.80D-5& 7.80D-6& 2.12D-5& 1.10D-5& 1.20D-5& 9.40D-6& 8.40D-5& 2.28D-5&3.5D-6\\
\    11&9.90D-6& 5.26D-4& 2.02D-5& 6.38D-6& 2.18D-5& 9.24D-6& 6.60D-6& 7.04D-6& 6.58D-5& 1.63D-5&3.9D-6\\
\    12&7.42D-6& 3.36D-4& 1.68D-5& 4.90D-6& 2.10D-5& 8.82D-6& 6.02D-6& 5.18D-6& 5.88D-5& 1.33D-5&2.4D-6\\
\    13&1.17D-5& 2.78D-4& 2.86D-5& 7.37D-6& 1.35D-5& 9.68D-6& 7.70D-6& 8.58D-6& 7.90D-5& 2.37D-5&1.7D-7\\
\    14&9.28D-6& 5.60D-5& 1.26D-5& 6.40D-6& 1.23D-5& 6.60D-6& 4.60D-6& 6.40D-6& 4.56D-5& 9.60D-6&3.2D-7\\
\    15&9.55D-6& 6.29D-5& 1.50D-5& 6.23D-6& 1.09D-5& 6.80D-6& 4.91D-6 &6.60D-6& 4.91D-5& 1.16D-5&3.4D-7\\
\    16&6.66D-6& 9.58D-5& 1.30D-5& 5.37D-6& 1.92D-5& 8.88D-6& 5.55D-6& 4.14D-6& 5.51D-5& 9.25D-6&5.2D-7\\
\    17&6.77D-5& 5.04D-4& 1.08D-4& 4.48D-5& 7.56D-5& 5.29D-5& 3.74D-5& 4.63D-5& 3.73D-4& 8.28D-5&4.0D-6\\
\    18&9.31D-5& 1.28D-3& 1.74D-4& 7.20D-5& 2.03D-4& 1.11D-4& 7.35D-5& 6.17D-5& 7.15D-4& 1.33D-4&1.1D-5\\
\    19&3.07D-4& 2.64D-3& 3.90D-4& 1.90D-4& 3.60D-4& 1.90D-4& 1.33D-4& 2.20D-4& 1.31D-3& 3.03D-4&2.0D-5\\
\    20&5.80D-4& 7.34D-3& 6.84D-4& 3.45D-4& 5.92D-4& 2.81D-4& 2.12D-4& 4.03D-4& 2.12D-3& 5.42D-4&7.0D-5\\
\    21&4.20D-4& 2.14D-3& 3.92D-4& 2.72D-4& 3.23D-4& 2.08D-4& 1.48D-4& 2.84D-4& 1.46D-3& 2.88D-4&1.2D-5\\
\    22&6.24D-4& 4.13D-3& 6.71D-4& 3.39D-4& 1.90D-4& 2.50D-4& 1.98D-4& 4.37D-4& 2.02D-3& 5.07D-4&3.0D-5\\
\    23&2.82D-4& 2.32D-3& 3.26D-4& 1.74D-4& 1.32D-4& 1.43D-4& 1.09D-4& 1.80D-4& 1.12D-3& 2.43D-4&1.4D-5\\
\    24&3.83D-4& 4.34D-3& 5.02D-4& 2.18D-4& 1.89D-4& 1.80D-4& 1.45D-4& 2.70D-4& 1.48D-3& 3.91D-4&2.6D-5\\
\    25&3.97D-4& 7.19D-3& 5.42D-4& 2.26D-4& 2.34D-4& 1.91D-4& 1.51D-4& 2.84D-4& 1.56D-3& 4.23D-4&4.6D-5\\
\    26&3.55D-4& 2.27D-3& 2.97D-4& 2.25D-4& 1.37D-4& 1.28D-4& 9.81D-5& 2.21D-4& 1.02D-3& 2.04D-4&9.0D-6\\
\    27&3.03D-4& 4.29D-3& 3.43D-4& 1.92D-4& 1.24D-4& 1.33D-4& 1.75D-4& 1.89D-4& 1.11D-3& 2.38D-4&1.3D-5\\
\    28&3.72D-4& 2.59D-3& 4.64D-4& 1.60D-4& 2.80D-5& 9.20D-5& 1.70D-4& 2.60D-4& 1.11D-3& 3.37D-4&8.2D-6\\
\    29&4.14D-4& 2.38D-3& 5.36D-4& 2.06D-4& 5.85D-5& 1.46D-4& 2.52D-4& 2.65D-4& 1.52D-3& 3.69D-4&5.4D-6\\
\    30&3.62D-4& 2.48D-3& 6.01D-4& 1.49D-4& 3.00D-5& 1.04D-4& 2.09D-4& 2.68D-4& 1.29D-3& 4.59D-4&8.1D-6\\
\    31&6.40D-4& 9.71D-3& 8.03D-4& 3.32D-4& 1.48D-4& 2.18D-4& 3.90D-4& 4.21D-4& 2.19D-3& 5.65D-4&1.3D-5\\
\    32&3.19D-4& 2.01D-2& 4.62D-4& 1.93D-4& 2.77D-4& 1.68D-4& 1.34D-4& 2.10D-4& 1.39D-3& 3.36D-4&3.8D-7\\
\    33&2.84D-4& 4.84D-3& 3.32D-4& 1.70D-4& 1.00D-4& 1.26D-4& 2.00D-4& 1.70D-4& 1.10D-3& 2.20D-4&6.4D-6\\
\    34&3.71D-4& 1.28D-2& 5.14D-4& 1.91D-4& 8.69D-5& 1.38D-4& 1.27D-4& 2.44D-4& 1.38D-3& 3.76D-4&2.3D-5\\
\    35&1.38D-4& 5.47D-2& 3.68D-4& 8.05D-5& 2.53D-4& 1.38D-4& 1.84D-5& 8.74D-5& 1.06D-3& 2.92D-4&7.6D-5\\
\    36&3.63D-6& 6.03D-5& 8.05D-6& 2.76D-6& 7.31D-6& 4.37D-6& 7.59D-6& 2.30D-6& 2.90D-5& 5.52D-6&2.8D-7\\
\    37&1.58D-5& 2.45D-4& 3.87D-5& 8.80D-6& 2.64D-5& 1.41D-5& 1.85D-5& 1.20D-5& 1.07D-4& 3.26D-5&1.8D-6\\
\    38&1.74D-5& 3.60D-4& 4.29D-5& 9.78D-6& 2.55D-5& 1.45D-5& 1.88D-5& 1.34D-5& 1.11D-4& 3.75D-5&2.6D-6\\
\    pl&1.23D-5& 3.44D-4& 5.32D-5& 6.13D-6& 1.91D-5& 1.64D-5& 1.61D-5& 1.07D-5& 1.03D-4& 5.63D-5&7.3D-6\\
\hline\\
\end{tabular}}
\end{table*}

%\flushleft
\begin{table*}
\caption{Optical lines (\erg)}
\tiny{
\begin{tabular}{l l l l l l l llllll l l l l l} \\ \hline \hline
 p & [OII]& [NeIII] &[SII] & HI & [OIII]& [AIV]& \Hb & [OIII]& [NII] & HeI& [OI] & \Ha & [NII] & [SII] & [SII] & [AIII] & [OII]\\
 \AA& 3727+&3869+&4070+&4340&4363&4713&4861&5007+&5755&5876&6300+&6563&6583+&6717&6731&7262&7322+\\ \hline
     1&  2.4D-5& 8.6D-7& 2.4D-8& 8.4D-7& 4.7D-7& 4.9D-8& 1.9D-6& 6.3D-6& 1.5D-7& 2.6D-7& 3.0D-8& 5.9D-6& 4.9D-6& 3.9D-7& 2.8D-7& 5.0D-7& 1.4D-6\\
     2&  2.4D-5& 8.8D-7& 2.7D-8& 5.8D-7& 4.7D-7& 5.1D-8& 1.3D-6& 6.5D-6& 1.5D-7& 1.8D-7& 2.3D-8& 4.2D-6& 3.9D-6& 3.6D-7& 2.4D-7& 4.2D-7& 1.5D-6\\
     3&  4.1D-5& 1.4D-6& 5.1D-8& 1.8D-6& 7.4D-7& 7.8D-8& 3.9D-6& 1.1D-5& 2.3D-7& 5.5D-7& 3.5D-8& 1.2D-5& 9.3D-6& 9.0D-7& 6.2D-7& 9.4D-7& 2.4D-6\\
     4&  9.6D-5& 2.3D-6& 1.6D-7& 6.1D-6& 1.2D-6& 1.2D-7& 1.4D-5& 1.7D-5& 5.4D-7& 1.8D-6& 9.5D-8& 4.1D-5& 3.1D-5& 3.6D-6& 2.6D-6& 2.2D-6& 5.1D-6\\
     5&  1.0D-4& 2.6D-6& 1.4D-7& 6.0D-6& 1.3D-6& 1.3D-7& 1.3D-5& 2.0D-5& 5.2D-7& 1.8D-6& 9.1D-8& 3.9D-5& 3.5D-5& 3.1D-6& 2.2D-6& 3.2D-6& 5.1D-6\\
     6&  8.6D-5& 2.5D-6& 1.2D-7& 3.7D-6& 1.2D-6& 1.2D-7& 8.3D-6& 1.9D-5& 5.0D-7& 1.2D-6& 6.6D-8& 2.5D-5& 2.3D-5& 2.3D-6& 1.6D-6& 2.2D-6& 4.9D-6\\
     7&  9.3D-5& 2.6D-6& 1.1D-7& 5.1D-6& 1.4D-6& 1.4D-7& 1.1D-5& 2.1D-5& 5.2D-7& 1.5D-6& 8.5D-8& 3.3D-5& 3.1D-5& 2.4D-6& 1.7D-6& 3.3D-6& 4.8D-6\\
     8&  6.7D-5& 2.2D-6& 8.9D-8& 3.4D-6& 1.1D-6& 7.4D-8& 7.4D-6& 1.9D-5& 3.4D-7& 1.1D-6& 5.2D-8& 2.2D-5& 2.2D-5& 1.6D-6& 1.2D-6& 2.9D-6& 3.2D-6\\
     9&  9.5D-5& 3.2D-6& 1.1D-7& 5.5D-6& 1.7D-6& 1.2D-7& 1.2D-5& 2.6D-5& 4.8D-7& 1.8D-6& 9.6D-8& 3.6D-5& 3.2D-5& 2.0D-6& 2.1D-6& 4.0D-6& 3.5D-6\\
    10&  1.2D-4& 3.3D-6& 1.9D-7& 9.7D-6& 1.7D-6& 1.3D-7& 2.1D-5& 2.7D-5& 6.1D-7& 3.1D-6& 1.5D-7& 6.3D-5& 5.4D-5& 4.0D-6& 2.8D-6& 6.1D-6& 5.0D-6\\
    11&  1.1D-4& 2.6D-6& 1.5D-7& 1.0D-5& 1.3D-6& 1.3D-7& 2.2D-5& 2.2D-5& 5.1D-7& 3.2D-6& 1.3D-7& 6.5D-5& 5.8D-5& 3.8D-6& 2.7D-6& 6.8D-6& 3.5D-6\\
    12&  7.4D-5& 2.5D-6& 1.1D-7& 6.3D-6& 1.2D-6& 1.2D-7& 1.4D-5& 2.2D-5& 3.6D-7& 2.1D-6& 9.6D-8& 4.1D-5& 3.4D-5& 2.5D-6& 3.2D-6& 4.4D-6& 2.9D-6\\
    13&  9.0D-5& 2.7D-6& 1.1D-7& 5.0D-6& 1.4D-6& 1.3D-7& 1.1D-5& 2.1D-5& 5.1D-7& 1.6D-6& 9.4D-8& 3.3D-5& 2.8D-5& 2.3D-6& 1.6D-6& 3.1D-6& 4.8D-6\\
    14&  1.6D-5& 1.8D-6& 3.9D-8& 8.9D-7& 9.3D-7& 9.3D-8& 2.0D-6& 1.4D-5& 2.0D-7& 2.9D-7& 2.0D-8& 6.2D-6& 6.2D-6& 4.5D-7& 3.2D-7& 9.1D-7& 2.0D-6\\
    15&  1.9D-5& 1.8D-6& 3.3D-8& 9.9D-7& 9.7D-7& 9.7D-8& 2.2D-6& 1.4D-5& 2.4D-7& 3.2D-7& 2.6D-8& 7.0D-6& 7.0D-6& 4.4D-7& 3.0D-7& 9.7D-7& 2.4D-6\\
    16&  3.6D-5& 2.3D-6& 3.0D-8& 1.7D-6& 1.3D-6& 1.3D-7& 3.7D-6& 1.8D-5& 1.9D-7& 5.6D-7& 3.7D-8& 1.1D-5& 9.0D-6& 5.7D-7& 3.9D-7& 1.3D-6& 2.0D-6\\
    17&  3.0D-4& 1.5D-5& 2.9D-7& 8.1D-6& 7.6D-6& 7.0D-7& 1.8D-5& 1.2D-4& 1.8D-6& 2.7D-6& 2.3D-7 &5.6D-5& 6.0D-5& 3.8D-6& 2.8D-6& 8.8D-6& 1.8D-5\\
    18&  5.3D-4& 3.2D-5& 6.4D-7& 2.2D-5& 1.6D-5& 1.6D-6& 4.9D-5& 2.5D-4& 2.9D-6& 7.2D-6& 4.9D-7& 1.5D-4& 1.4D-4& 9.6D-6& 7.2D-6& 2.1D-5& 3.0D-5\\
    19&  1.1D-3& 5.5D-5& 1.4D-6& 4.5D-5& 2.7D-5& 2.6D-6& 9.9D-5& 4.5D-4& 6.9D-6& 1.5D-5& 9.7D-7& 3.0D-4& 2.9D-4& 1.6D-5& 1.4D-5& 4.3D-5& 6.8D-5\\
    20&  2.3D-3& 7.8D-5& 5.2D-6& 1.3D-4& 4.1D-5& 4.1D-6& 2.9D-4& 6.5D-4& 1.4D-5& 4.2D-5& 2.4D-6& 8.7D-4& 9.4D-4& 4.7D-5& 5.2D-5& 1.2D-4& 1.4D-4\\
    21&  1.1D-3& 6.1D-5& 1.1D-6& 3.7D-5& 3.1D-5& 2.8D-6& 8.1D-5& 5.0D-4& 6.2D-6& 1.2D-5& 6.2D-7& 2.5D-4& 2.3D-4& 1.1D-5& 9.7D-6& 3.9D-5& 7.1D-5\\
    22&  1.8D-3& 8.0D-5& 3.1D-6& 7.2D-5& 3.7D-5& 3.0D-6& 1.6D-4& 6.6D-4& 1.1D-5& 2.3D-5& 1.4D-6& 4.8D-4& 4.6D-4& 2.7D-5& 2.7D-5& 7.3D-5& 1.3D-4\\
    23&  9.7D-4& 4.1D-5& 1.2D-6& 4.1D-5& 2.1D-5& 1.8D-6& 9.0D-5& 3.3D-4& 5.4D-6& 1.3D-5& 7.0D-7& 2.7D-4& 2.5D-4& 1.6D-5& 1.3D-5& 3.6D-5& 5.8D-5\\
    24&  1.6D-3& 5.4D-5& 2.6D-6& 8.1D-5& 2.7D-5& 2.5D-6& 1.7D-4& 4.5D-4& 8.8D-6& 2.6D-5& 1.2D-6& 5.3D-4& 4.8D-4& 3.0D-5& 2.7D-5& 6.5D-5& 9.6D-5\\
    25&  1.9D-3& 5.7D-5& 4.1D-6& 1.3D-4& 2.8D-5& 2.6D-6& 2.9D-4& 4.8D-4& 1.1D-5& 4.3D-5& 1.9D-6& 8.8D-4& 7.7D-4& 5.2D-5& 5.0D-5& 9.7D-5& 1.1D-4\\
    26&  8.4D-4& 3.5D-5& 1.0D-6& 4.2D-5& 1.8D-5& 1.8D-6& 9.1D-5& 2.8D-4& 4.5D-6& 1.4D-5& 5.2D-7& 2.7D-4& 2.3D-4& 1.4D-5& 1.1D-5& 3.1D-5& 5.2D-5\\
    27&  1.0D-3& 3.7D-5& 1.7D-6& 8.1D-5& 1.9D-5& 1.7D-6& 1.7D-4& 2.9D-4& 5.2D-6& 2.6D-5& 6.6D-7& 5.3D-4& 4.0D-4& 2.8D-5& 2.4D-5& 4.4D-5& 5.9D-5\\
    28&  1.3D-3& 3.1D-5& 2.0D-6& 4.8D-5& 1.5D-5& 1.3D-6& 1.1D-4& 2.3D-4& 7.4D-6& 1.6D-5& 7.0D-7& 3.3D-4& 2.9D-4& 2.3D-5& 2.1D-5& 3.5D-5& 8.8D-5\\
    29&  1.3D-3& 4.4D-5& 2.3D-6& 4.0D-5& 2.2D-5& 1.8D-6& 9.0D-5& 3.2D-4& 7.7D-6& 1.4D-5& 6.2D-7& 2.8D-4& 2.4D-4& 2.3D-5& 2.0D-5& 3.2D-5& 9.5D-5\\
    30&  1.4D-3& 3.2D-5& 3.4D-6& 4.1D-5& 1.5D-5& 1.3D-6& 9.1D-5& 2.4D-4& 9.1D-6& 1.3D-5& 7.6D-7& 2.8D-4& 2.8D-4& 2.9D-5& 2.8D-5& 3.4D-5& 1.1D-4\\
    31&  1.9D-3& 6.4D-5& 9.0D-6& 1.8D-4& 3.2D-5& 2.2D-6& 3.9D-4& 4.7D-4& 1.2D-5& 5.8D-5& 1.0D-6& 1.2D-3& 7.4D-4& 8.6D-5& 1.1D-4& 6.2D-5& 1.6D-4\\
    32&  2.0D-3& 4.7D-5& 1.0D-5& 3.9D-4& 2.4D-5& 1.8D-6& 8.4D-4& 3.5D-4& 1.0D-5& 1.2D-4& 1.7D-6& 2.5D-3& 1.5D-3& 2.0D-4& 1.9D-4& 8.4D-5& 8.4D-5\\
    33&  9.6D-4& 3.5D-5& 1.8D-6& 9.2D-5& 1.8D-5& 1.2D-6& 2.0D-4& 2.7D-4& 5.2D-6& 3.1D-5& 4.4D-7& 6.0D-4& 3.8D-4& 3.0D-5& 2.6D-5& 4.2D-5& 5.8D-5\\
    34&  1.7D-3& 4.0D-5& 7.4D-6& 2.4D-4& 2.0D-5& 1.4D-6& 5.3D-4& 3.0D-4& 9.5D-6& 7.5D-5& 1.1D-6& 1.6D-3& 1.0D-3& 1.2D-4& 1.2D-4& 6.4D-5& 9.8D-5\\
    35&  2.3D-2& 3.5D-5& 3.9D-5& 1.1D-3& 1.8D-5& 1.9D-6& 2.3D-3& 2.5D-4& 9.2D-6& 7.4D-5& 3.9D-6& 6.9D-3& 2.9D-3& 1.0D-3& 9.4D-4& 1.8D-5& 6.9D-5\\
    36&  2.0D-5& 1.2D-6& 1.8D-8& 1.0D-6& 6.2D-7& 6.4D-8& 2.3D-6& 8.7D-6& 1.1D-7& 3.4D-7& 3.2D-8& 7.1D-6& 5.0D-6& 3.8D-7& 2.6D-7& 5.8D-7& 1.2D-6\\
    37&  1.1D-4& 4.1D-6& 1.1D-7& 4.0D-6& 2.0D-6& 1.4D-7& 8.8D-6& 3.0D-5& 6.7D-7& 1.3D-6& 1.7D-7& 2.8D-5& 2.6D-5& 1.8D-6& 1.3D-6& 3.0D-6& 6.3D-6\\
    38&  1.3D-4& 4.2D-6& 1.6D-7& 6.0D-6& 2.1D-6& 2.0D-7& 1.3D-5& 3.0D-5& 7.8D-7& 1.9D-6& 3.1D-7& 4.2D-5& 3.6D-5& 3.0D-6& 2.2D-6& 3.6D-6& 7.1D-6\\ 
    pl&  2.4D-4& 4.3D-6& 2.0D-6& 2.1D-6& 2.4D-6& 3.1D-7& 3.1D-6& 3.3D-5& 1.1D-6& 1.4D-6& 9.4D-7& 2.2D-5& 2.1D-5& 1.2D-5& 9.0D-6& 1.7D-6& 1.2D-5\\
\hline
\end{tabular}
}
\end{table*}

\begin{table*}
\begin{center}
\caption{IR lines (\erg)}
\tiny{
\begin{tabular}{l l l l l l llllll l l l l l} \\ \hline \hline
\ position& [SIII]&[CI]&[SII]&HeI&HI&HI&[AII]&[AIII]&[SIV]&[NeII]&[NeIII]&[SIII]&[AIII]\\
\ \mum & 0.94&0.98&1.03&1.08&1.28&2.34&6.98&8.99&10.5&12.8&15.6&18.7&21.8\\ \hline
\     1& 9.5D-7& 5.6D-8& 2.2D-8& 3.7D-7& 3.2D-7& 7.1D-7& 9.9D-8& 3.7D-7& 3.0D-8& 1.0D-6& 1.2D-7& 3.7D-7& 2.8D-8\\
\     2& 8.3D-7& 5.1D-8& 2.4D-8& 2.6D-7& 2.2D-7& 4.9D-7& 7.3D-8& 2.4D-7& 3.7D-8& 6.9D-7& 9.1D-8& 2.8D-7& 1.7D-8\\
\     3& 2.7D-6& 5.5D-8& 4.7D-8& 8.2D-7& 6.7D-7& 1.5D-6& 2.1D-7& 7.4D-7& 7.0D-8& 2.2D-6& 1.8D-7& 1.1D-6& 5.6D-8\\
\     4& 7.7D-6& 1.3D-7& 1.5D-7& 2.6D-6& 2.3D-6& 5.2D-6& 1.1D-6& 1.8D-6& 1.2D-7& 7.6D-6& 2.6D-7& 3.5D-6& 1.3D-7\\
\     5& 1.1D-5& 1.1D-7& 1.3D-7& 2.7D-6& 2.2D-6& 4.9D-6& 6.5D-7& 2.9D-6& 1.3D-7& 7.7D-6& 3.8D-7& 4.4D-6& 2.1D-7\\
\     6& 7.6D-6& 8.3D-8& 1.1D-7& 1.7D-6& 1.4D-6& 3.2D-6& 4.2D-7& 1.7D-6& 1.4D-7& 4.8D-6& 3.0D-7& 3.1D-6& 1.3D-7\\
\     7& 1.0D-5& 9.8D-8& 1.1D-7& 2.2D-6& 1.9D-6& 4.2D-6& 4.2D-7& 2.9D-6& 1.4D-7& 6.5D-6& 5.0D-7& 4.1D-6& 2.1D-7\\
\     8& 1.1D-5& 5.2D-8& 8.2D-8& 1.6D-6& 1.3D-6& 2.8D-6& 1.6D-7& 2.4D-6& 1.7D-7& 4.5D-6& 6.9D-7& 4.0D-6& 1.8D-7\\
\     9& 1.0D-5& 1.1D-7& 9.5D-8& 2.7D-6& 2.1D-6& 4.6D-6& 3.3D-7& 3.7D-6& 1.6D-7& 7.1D-6& 9.8D-7& 4.2D-6& 2.7D-7\\
\    10& 1.8D-5& 1.7D-7& 1.6D-7& 4.3D-6& 3.5D-6& 7.9D-6& 6.6D-7& 5.8D-6& 1.9D-7& 1.2D-5& 1.2D-6& 7.9D-6& 4.1D-7\\
\    11& 2.2D-5& 1.3D-7& 1.5D-7& 4.6D-6& 3.7D-6& 8.4D-6& 5.9D-7& 6.6D-6& 2.0D-7& 1.3D-5& 1.5D-6& 9.0D-6& 4.8D-7\\
\    12& 1.5D-5& 9.6D-8& 9.6D-8& 3.0D-6& 2.3D-6& 5.2D-6& 2.7D-7& 4.4D-6& 1.9D-7& 7.8D-6& 1.7D-6& 6.3D-6& 3.2D-7\\
\    13& 9.2D-6& 1.2D-7& 1.1D-7& 2.3D-6& 1.9D-6& 4.1D-6& 4.0D-7& 2.8D-6& 1.4D-7& 6.3D-6& 6.1D-7& 3.8D-6& 2.1D-7\\
\    14& 2.6D-6& 3.0D-8& 3.0D-8& 4.3D-7& 3.3D-7& 7.3D-7& 3.9D-8& 6.1D-7& 9.7D-8& 1.1D-6& 3.0D-7& 8.9D-7& 4.5D-8\\
\    15& 2.1D-6& 4.0D-8& 2.9D-8& 4.8D-7& 3.7D-7& 8.1D-7& 5.7D-8& 6.6D-7& 7.7D-8& 1.3D-6& 2.9D-7& 7.3D-7& 4.8D-8\\
\    16& 3.0D-6& 4.4D-8& 3.0D-8& 8.1D-7& 6.3D-7& 1.4D-6& 8.1D-8& 1.1D-6& 1.0D-7& 2.0D-6& 6.2D-7& 1.2D-6& 8.1D-8\\
\    17& 2.2D-5& 3.1D-7& 2.7D-7& 4.0D-6& 3.1D-6& 6.7D-6& 3.4D-7& 5.9D-6& 7.0D-7& 1.0D-5& 3.5D-6& 7.6D-6& 4.3D-7\\
\    18& 6.4D-5& 5.9D-7& 5.9D-7& 1.1D-5& 8.3D-6& 1.8D-5& 8.3D-7& 1.6D-5& 1.8D-6& 2.7D-5& 1.1D-5& 2.4D-5& 1.2D-6\\
\    19& 1.2D-4& 1.2D-6& 1.2D-6& 2.4D-5& 1.7D-5& 3.7D-5& 1.7D-6& 3.3D-5& 3.0D-6& 5.5D-5& 2.1D-5& 4.7D-5& 2.4D-6\\
\    20& 3.8D-4& 2.3D-6& 4.6D-6& 7.5D-5& 4.9D-5& 1.1D-4& 7.5D-6& 9.0D-5& 4.9D-6& 1.8D-4& 2.2D-5& 1.5D-4& 6.7D-6\\
\    21& 1.2D-4& 6.9D-7& 1.1D-6& 2.0D-5& 1.4D-5& 3.0D-5& 1.1D-6& 2.8D-5& 3.9D-6& 4.6D-5& 1.8D-5& 4.5D-5& 2.0D-6\\
\    22& 2.7D-4& 1.5D-6& 2.8D-6& 4.1D-5& 2.6D-5& 5.8D-5& 2.2D-6& 5.3D-5& 5.9D-6& 8.9D-5& 3.1D-5& 1.1D-4& 3.9D-6\\
\    23& 1.2D-4& 7.7D-7& 1.1D-6& 2.1D-5& 1.5D-5& 3.3D-5& 1.7D-6& 2.9D-5& 2.6D-6& 5.2D-5& 1.1D-5& 4.6D-5& 2.1D-6\\
\    24& 2.3D-4& 1.3D-6& 2.3D-6& 4.2D-5& 3.0D-5& 6.5D-5& 3.5D-6& 5.5D-5& 3.8D-6& 1.0D-4& 1.7D-5& 1.0D-4& 4.0D-6\\
\    25& 3.8D-4& 2.0D-6& 3.8D-6& 7.0D-5& 4.9D-5& 1.1D-4& 7.0D-6& 9.0D-5& 4.6D-6& 1.7D-4& 2.3D-5& 1.7D-4& 6.4D-6\\
\    26& 1.0D-4 &5.0D-7 &8.6D-7 &2.1D-5 &1.5D-5 &3.5D-5& 2.5D-6& 2.6D-5& 2.1D-6& 5.3D-5& 5.6D-6& 4.3D-5& 1.9D-6\\
\    27& 1.9D-4 &6.1D-7 &1.5D-6 &3.8D-5 &3.0D-5 &6.6D-5 &6.1D-6& 4.5D-5& 2.5D-6& 1.0D-4& 6.0D-6& 9.1D-5& 3.3D-6\\
\    28& 1.6D-4 &7.0D-7 &1.8D-6 &2.5D-5 &1.8D-5 &4.1D-5 &3.9D-6 &2.8D-5& 2.5D-6& 6.2D-5& 4.0D-6& 6.7D-5& 2.0D-6\\
\    29& 1.8D-4 &5.6D-7 &2.0D-6 &2.1D-5 &1.5D-5 &3.4D-5 &2.7D-6 &2.4D-5& 4.3D-6& 5.2D-5& 4.8D-6& 7.7D-5& 1.8D-6\\
\    30& 1.9D-4 &8.6D-7 &3.0D-6 &2.2D-5 &1.5D-5 &3.5D-5 &3.5D-6 &2.3D-5& 3.4D-6& 5.3D-5& 3.6D-6& 7.6D-5& 1.6D-6\\
\    31& 6.1D-4 &7.4D-7 &8.2D-6 &9.4D-5 &6.6D-5 &1.5D-4 &2.5D-5 &6.4D-5& 7.4D-6& 2.2D-4& 6.2D-6& 3.5D-4& 4.7D-6\\
\    32& 8.1D-4 &1.5D-6 &9.2D-6 &1.8D-4 &1.5D-4 &3.3D-4 &6.4D-5 &1.1D-4& 4.4D-6& 4.7D-4& 8.4D-6& 5.0D-4& 7.6D-6\\
\    33& 2.6D-4 &3.0D-7 &1.5D-6 &4.5D-5 &3.5D-5 &7.7D-5 &7.2D-6 &5.0D-5& 3.2D-6& 1.1D-4& 4.4D-6& 1.4D-4& 3.6D-6\\
\    34& 5.3D-4 &9.5D-7 &6.9D-6 &1.2D-4 &9.3D-5 &2.1D-4 &4.0D-5& 7.2D-5& 3.4D-6& 3.0D-4& 4.7D-6& 3.1D-4& 5.1D-6\\
\    35& 1.2D-3 &3.5D-6 &3.5D-5 &1.1D-4 &4.1D-4 &9.1D-4 &2.6D-4& 1.6D-5& 3.9D-6& 1.2D-3& 6.2D-6& 9.0D-4& 1.2D-6\\
\    36& 1.1D-6 &4.6D-8 &1.7D-8 &4.8D-7 &4.0D-7 &8.9D-7 &9.9D-8& 5.3D-7& 4.1D-8& 1.3D-6& 2.1D-7& 5.1D-7& 3.9D-8\\
\    37& 5.5D-6 &2.6D-7 &1.0D-7 &1.8D-6 &1.5D-6 &3.3D-6 &3.3D-7& 2.3D-6& 1.4D-7& 4.9D-6& 1.0D-6& 2.1D-6& 1.7D-7\\
\    38& 7.0D-6 &4.6D-7 &1.5D-7 &2.8D-6 &2.3D-6 &5.1D-6 &6.4D-7& 3.1D-6& 1.3D-7& 7.6D-6& 1.1D-6& 2.8D-6& 2.2D-7\\
\    pl& 1.1D-5 & 1.4D-6&1.4D-6 &1.5D-6 &8.5D-7 &2.0D-6 &8.5D-7& 2.9D-7& 6.1D-7& 3.9D-6& 1.1D-6& 1.7D-6& 2.2D-8\\ 
\hline
\end{tabular}}
\end{center}
\end{table*}

\begin{table*}
\caption{IR lines (\erg)}
\tiny{
\begin{tabular}{l l l l l l llllll l l l l l} \\ \hline \hline
\ position
&[FeIII]&[OIV]&[FeII]&[SIII]&[SiII]&[NeIII]&[OIII]&[NIII]&[OI]&[OIII]&[NII]&[CII]&[NII]\\
\ \mum  
&22.9&25.9&26.0&33.6&34.8&36.1&51.7&57.3&63.1&87.3&121.5&156.&203.3\\ \hline
\     1
& 1.7D-8& 3.4D-8& 8.2D-8& 6.9D-7& 2.0D-6& 1.1D-8& 1.6D-7& 2.4D-7& 3.7D-8& 2.6D-7& 5.2D-7& 2.9D-6& 5.4D-7\\
\     2
& 1.8D-8& 3.8D-8& 9.2D-8& 5.2D-7& 1.4D-6& 7.7D-9& 1.3D-7& 1.8D-7& 2.6D-8& 2.2D-7& 3.5D-7& 2.0D-6& 3.6D-7\\
\     3
& 7.8D-8& 3.9D-8& 1.8D-7& 2.1D-6& 4.3D-6& 1.6D-8& 3.0D-7& 5.4D-7& 3.9D-8& 5.0D-7& 1.1D-6& 5.1D-6& 1.0D-6\\
\     4
& 1.9D-7& 3.8D-8& 3.8D-7& 6.3D-6& 1.2D-5& 2.7D-8& 4.5D-7& 8.9D-7& 1.1D-7& 7.0D-7& 3.5D-6& 9.9D-6& 2.1D-6\\
\     5
& 3.6D-7& 5.7D-8& 3.5D-7& 8.1D-6& 1.1D-5& 3.9D-8& 8.3D-7& 1.9D-6& 7.8D-8& 1.3D-6& 3.2D-6& 8.7D-6& 2.0D-6\\
\     6
& 3.3D-7& 4.6D-8& 3.8D-7& 5.6D-6& 8.1D-6& 2.5D-8& 6.0D-7& 1.2D-6& 5.8D-8& 9.1D-7& 2.2D-6& 6.0D-6& 1.3D-6\\
\     7
& 4.1D-7& 5.9D-8& 3.5D-7& 7.4D-6& 9.6D-6& 4.4D-8& 1.1D-6& 2.4D-6& 6.6D-8& 1.8D-6& 2.8D-6& 7.5D-6& 1.7D-6\\
\     8
& 4.2D-7& 6.1D-8& 1.9D-7& 7.4D-6& 6.2D-6& 6.0D-8& 1.7D-6& 3.5D-6& 3.7D-8& 2.8D-6& 1.8D-6& 6.1D-6& 1.4D-6\\
\     9
& 6.5D-7& 1.1D-7& 5.3D-7& 7.7D-6& 1.1D-5& 8.6D-8& 2.1D-6& 3.9D-6& 8.6D-8& 3.4D-6& 3.0D-6& 8.5D-6& 2.0D-6\\
\    10
& 5.2D-7& 9.3D-8& 4.3D-7& 1.4D-5& 1.5D-5& 1.0D-7& 2.7D-6& 5.4D-6& 1.4D-7& 4.3D-6& 5.2D-6& 1.5D-5& 3.5D-6\\
\    11
& 7.9D-7& 9.0D-8& 4.6D-7& 1.7D-5& 1.6D-5& 1.3D-7& 3.7D-6& 7.7D-6& 1.3D-7& 6.2D-6& 5.7D-6& 2.2D-5& 4.8D-6\\
\    12
& 5.2D-7& 8.8D-8& 3.0D-7& 1.2D-5& 9.0D-6& 1.4D-7& 3.6D-6& 6.4D-6& 8.2D-8& 6.0D-6& 3.3D-6& 1.4D-5& 3.2D-6\\
\    13
& 3.8D-7& 5.7D-8& 4.8D-7& 7.1D-6& 6.9D-6& 5.4D-8& 1.9D-6& 2.3D-6& 8.7D-8& 1.9D-6& 2.7D-6& 8.4D-6& 1.9D-6\\
\    14
& 1.9D-7& 5.1D-8& 1.4D-7& 1.7D-6& 2.1D-6& 2.8D-8& 5.9D-7& 9.9D-7& 1.2D-8& 1.0D-6& 4.7D-7& 2.5D-6& 5.1D-7\\
\   15
& 2.0D-7& 4.6D-8& 2.2D-7& 1.4D-6& 2.4D-6& 2.6D-8& 5.3D-7& 8.6D-7& 1.8D-8& 8.8D-7& 5.5D-7& 2.7D-6& 5.5D-7\\
\    16
& 2.8D-7& 8.1D-8& 3.3D-7& 2.3D-6& 3.3D-6& 5.5D-8& 1.0D-6& 1.5D-6& 3.3D-8& 1.7D-6& 9.2D-7& 5.3D-6& 1.0D-6\\
\    17
& 1.4D-6& 3.2D-7& 1.1D-6& 1.3D-5& 1.5D-5& 3.1D-7& 6.3D-6& 8.1D-6& 1.4D-7& 8.8D-6& 3.3D-6& 5.4D-6& 1.3D-6\\
\    18
& 3.7D-6& 8.8D-7& 2.7D-6& 4.0D-5& 3.2D-5& 9.8D-7& 2.0D-5& 2.2D-5& 3.4D-7& 2.5D-5& 7.7D-6& 1.0D-5& 2.5D-6\\
\    19
& 1.2D-5& 1.5D-6& 7.5D-6& 6.4D-5& 7.4D-5& 1.8D-6& 3.9D-5& 3.2D-5& 5.9D-7& 3.3D-5& 8.4D-6& 7.8D-6& 1.6D-6\\
\    20
& 2.1D-5& 2.3D-6& 1.0D-5& 1.4D-4& 1.2D-4& 2.0D-6& 4.9D-5& 3.5D-5& 1.2D-6& 4.9D-5& 1.2D-5& 9.6D-6& 1.7D-6\\
\    21
& 1.1D-5& 1.4D-6& 3.3D-6& 6.1D-5& 5.4D-5& 1.5D-6& 4.3D-5& 3.6D-5& 2.4D-7& 3.6D-5& 6.1D-6& 5.6D-6& 1.2D-6\\
\    22
& 8.4D-6& 8.3D-7& 3.4D-6& 1.1D-4& 7.3D-5& 2.7D-6& 6.4D-5& 3.9D-5& 6.2D-7& 6.4D-5& 7.0D-6& 5.5D-6& 1.0D-6\\
\    23
& 5.8D-6& 5.9D-7& 3.0D-6& 6.9D-5& 6.4D-5& 9.9D-7& 2.7D-5& 3.1D-5& 4.5D-7& 2.7D-5& 1.1D-5& 1.0D-5& 2.3D-6\\
\    24
& 1.5D-5& 8.2D-7& 7.0D-6& 1.2D-4& 9.4D-5& 1.4D-6& 4.2D-5& 3.8D-5& 7.0D-7& 3.0D-5& 1.2D-5& 1.0D-5& 2.1D-6\\
\    25
& 1.5D-5& 9.9D-7& 8.4D-6& 2.0D-4& 1.5D-4& 2.0D-6& 5.5D-5& 4.8D-5& 1.4D-6& 3.5D-5& 1.8D-5& 1.5D-5& 2.9D-6\\
\    26
& 8.6D-6& 6.0D-7& 3.8D-6& 6.4D-5& 7.6D-5& 4.5D-7& 1.6D-5& 2.2D-5& 2.7D-7& 1.5D-5& 1.0D-5& 1.1D-5& 2.3D-6\\
\    27
& 6.5D-6& 5.2D-7& 2.8D-6& 1.3D-4& 1.0D-4& 5.2D-7& 1.8D-5& 2.9D-5& 5.2D-7& 1.8D-5& 1.9D-5& 1.9D-5& 4.0D-6\\
\    28
& 3.7D-6& 1.1D-7& 1.8D-6& 8.5D-5& 5.3D-5& 3.2D-7& 1.1D-5& 1.5D-5& 3.2D-7& 1.1D-5& 8.6D-6& 7.5D-6& 1.5D-6\\
\    29
& 2.2D-6& 2.3D-7& 9.9D-7& 1.0D-4& 3.6D-5& 4.5D-7& 1.4D-5& 1.2D-5& 1.8D-7& 1.2D-5& 7.5D-6& 6.8D-6& 1.4D-6\\
\    30
& 1.5D-6& 1.2D-7& 9.1D-7& 8.6D-5& 3.3D-5& 3.6D-7& 9.1D-6& 1.1D-5& 2.7D-7& 6.6D-6& 5.5D-6& 4.8D-6& 8.6D-7\\
\    31
& 3.5D-6& 6.2D-7& 1.6D-6& 2.7D-4& 7.4D-5& 3.9D-7& 1.2D-5& 1.2D-5& 3.9D-7& 7.4D-6& 1.2D-5& 9.8D-6& 1.6D-6\\
\    32
& 1.6D-5& 1.2D-6& 1.1D-5& 6.0D-4& 3.6D-4& 7.6D-7& 1.8D-5& 2.5D-5& 1.7D-6& 1.3D-5& 6.3D-5& 6.0D-5& 1.0D-5\\
\    33
& 5.0D-6& 4.2D-7& 1.4D-6& 1.9D-4& 6.2D-5& 4.0D-7& 1.5D-5& 3.0D-5& 2.0D-7& 1.4D-5& 1.9D-5& 1.8D-5& 3.8D-6\\
\    34
& 4.7D-6& 3.6D-7& 2.6D-6& 3.3D-4& 1.6D-4& 4.2D-7& 1.1D-5& 1.5D-5& 1.1D-6& 7.4D-6& 3.1D-5& 2.7D-5& 4.5D-6\\
\    35
& 1.5D-5& 9.7D-7& 3.0D-5& 1.1D-3& 7.8D-4& 4.6D-7& 6.9D-6& 4.6D-6& 9.2D-6& 6.9D-6& 1.8D-4& 1.9D-4& 3.0D-5\\
\    36
& 3.0D-8& 3.2D-8& 1.3D-7& 9.4D-7& 2.3D-6& 1.8D-8& 2.5D-7& 3.7D-7& 4.4D-8& 4.5D-7& 6.3D-7& 4.8D-6& 8.0D-7\\
\    37
& 1.4D-7& 1.1D-7& 4.8D-7& 3.8D-6& 8.1D-6& 8.8D-8& 1.1D-6& 1.5D-6& 1.8D-7& 1.8D-6& 2.2D-6& 5.6D-6& 1.3D-6\\
\    38
& 1.0D-7& 1.0D-7& 5.2D-7& 5.1D-6& 1.3D-5& 5.4D-8& 1.1D-6& 1.4D-6& 3.8D-7& 1.6D-6& 3.2D-6& 7.2D-6& 1.6D-6\\
\    pl
& 3.1D-8& 1.1D-7& 6.6D-7& 3.1D-6& 9.4D-6& 9.1D-8& 5.5D-7& 5.2D-7& 6.8D-6& 8.8D-7& 9.7D-7& 1.2D-5& 6.4D-7\\ 
\hline
\end{tabular}}
\end{table*}

\begin{figure*}

\begin{center}
\includegraphics[width=0.49\textwidth]{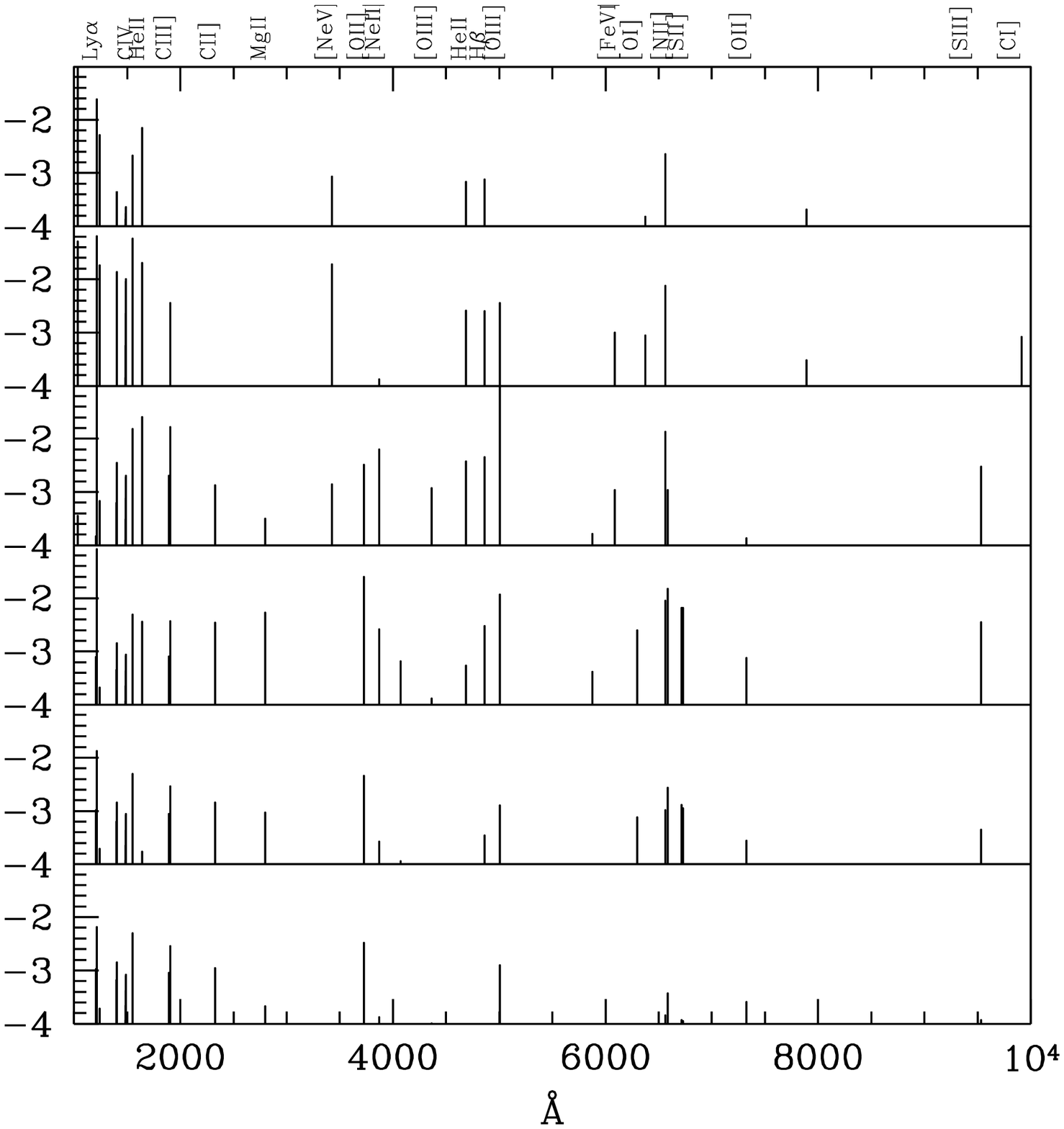}
\includegraphics[width=0.49\textwidth]{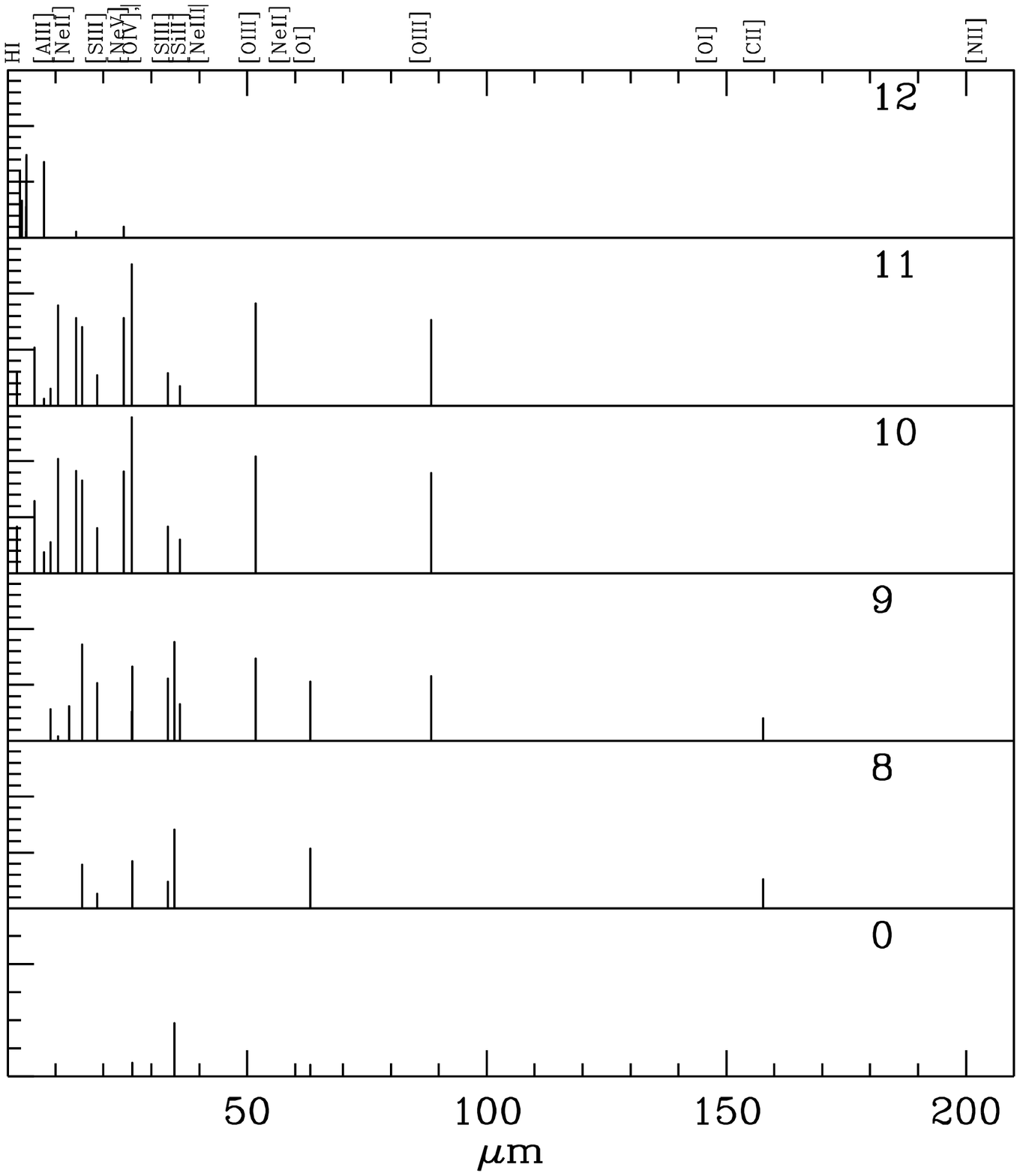}
\includegraphics[width=0.49\textwidth]{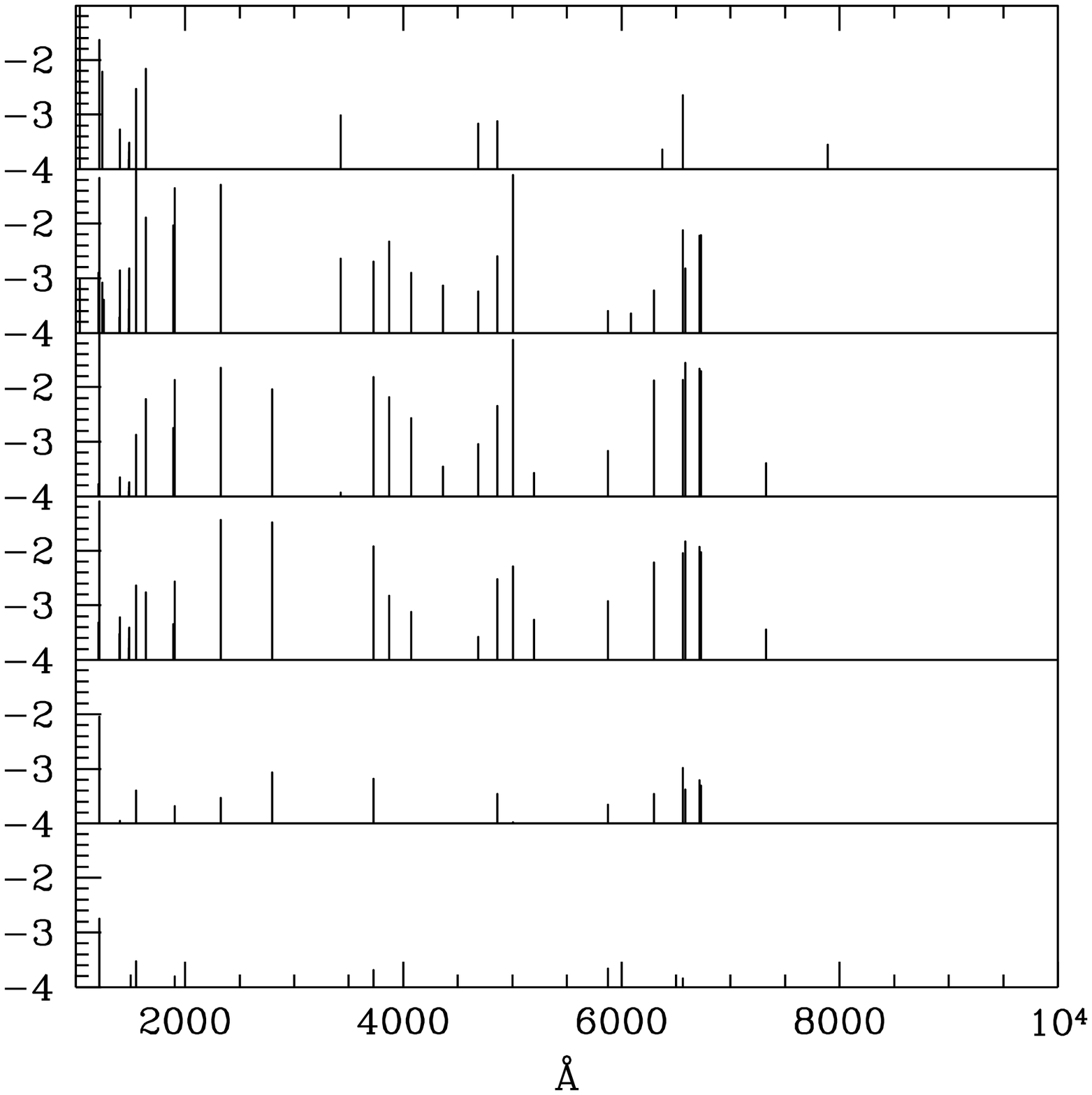}
\includegraphics[width=0.49\textwidth]{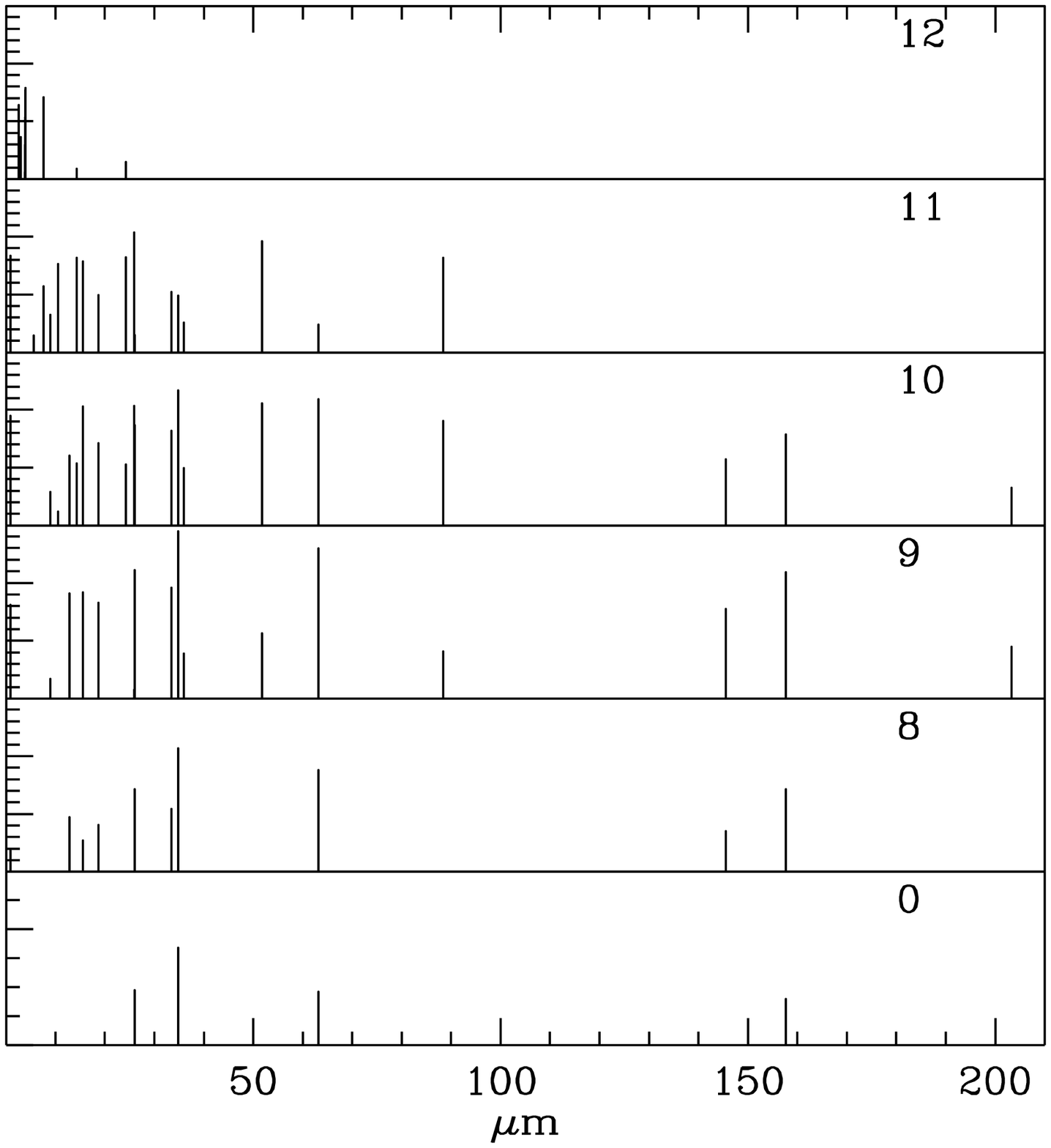}
\end{center}
\caption{Model calculations of line fluxes adopting a pl flux (Contini \& Viegas 2001a);
see text}
\end{figure*}
\begin{figure*}
\begin{center}
\includegraphics[width=0.49\textwidth]{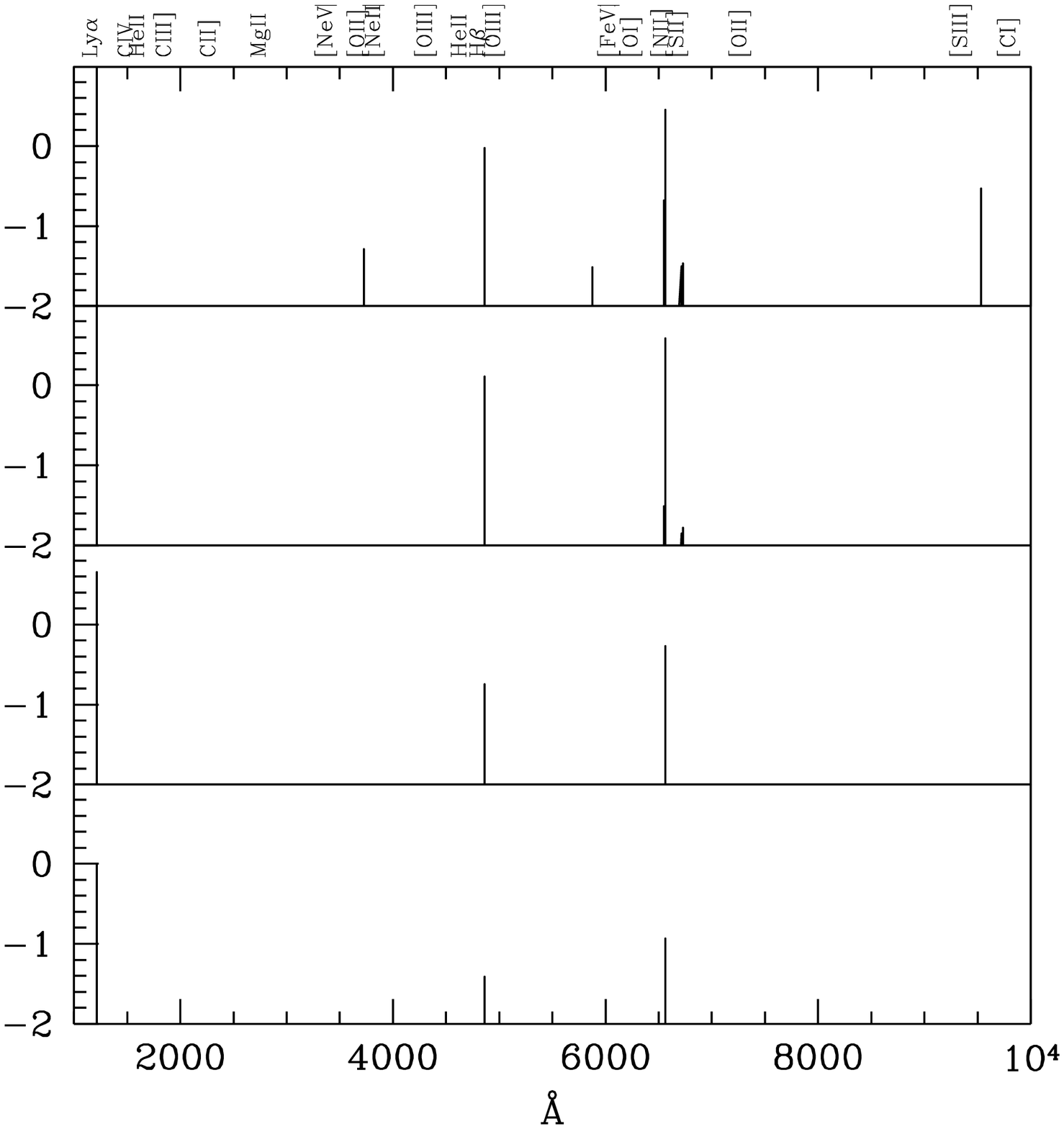}
\includegraphics[width=0.49\textwidth]{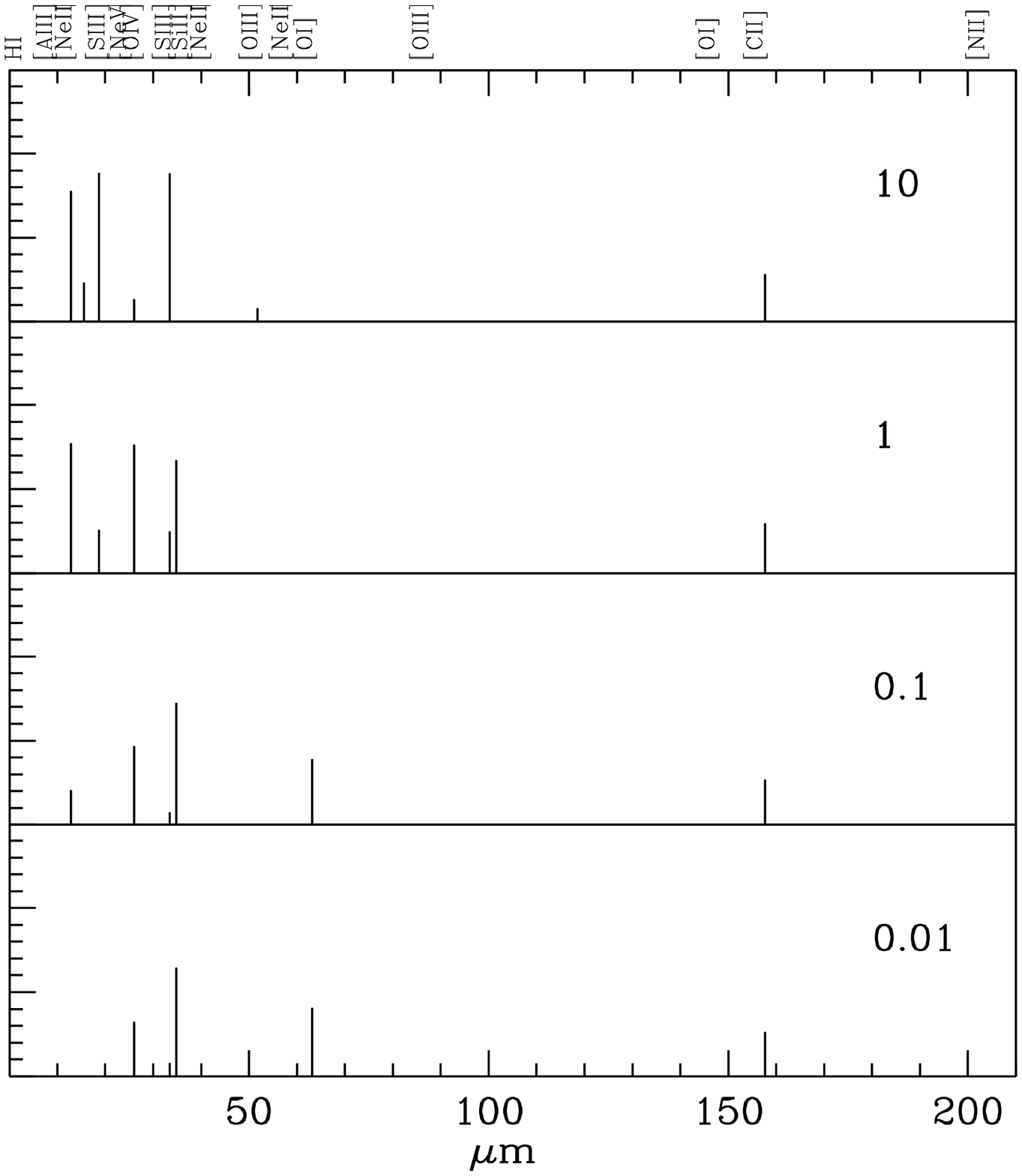}
\includegraphics[width=0.49\textwidth]{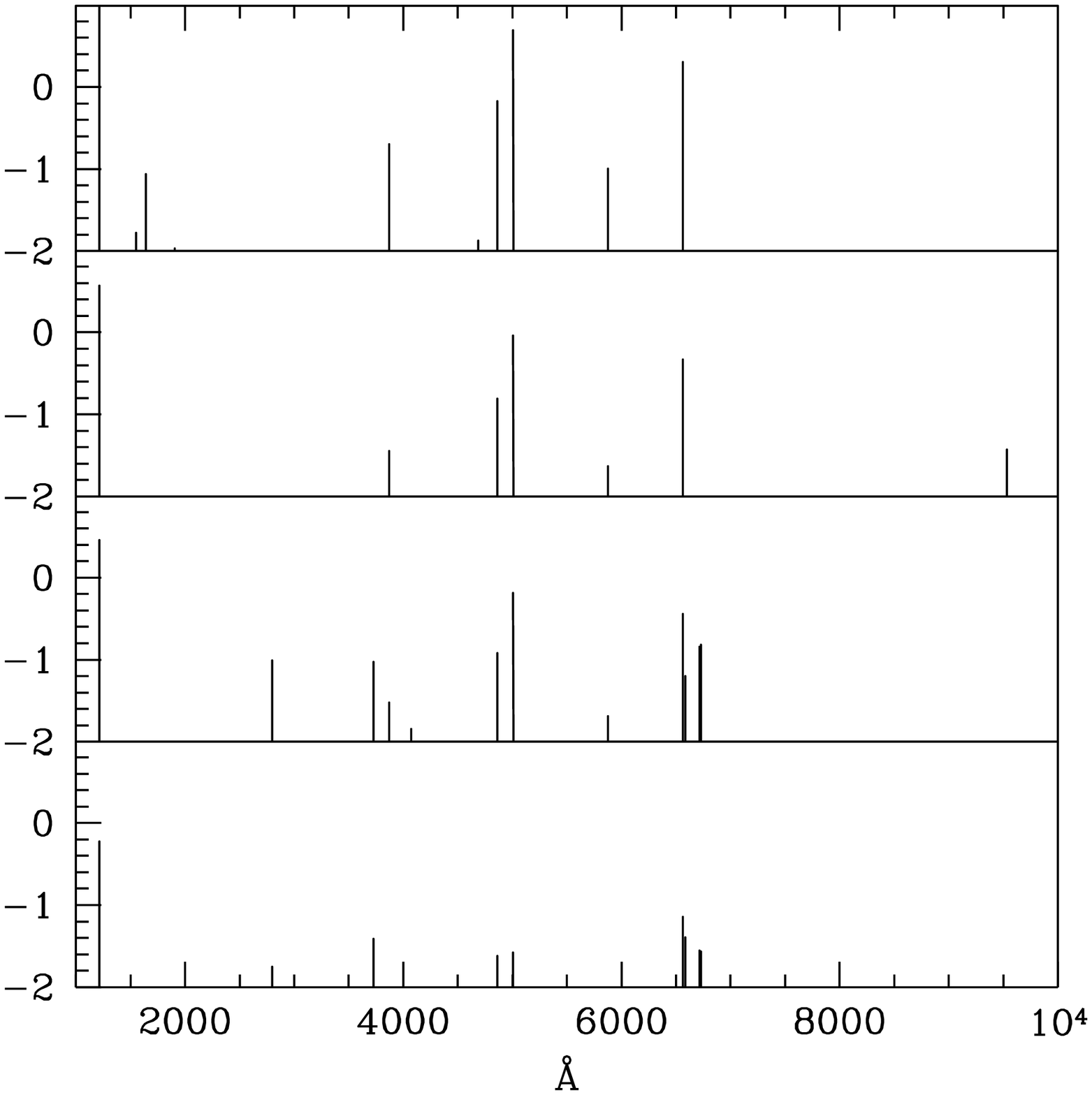}
\includegraphics[width=0.49\textwidth]{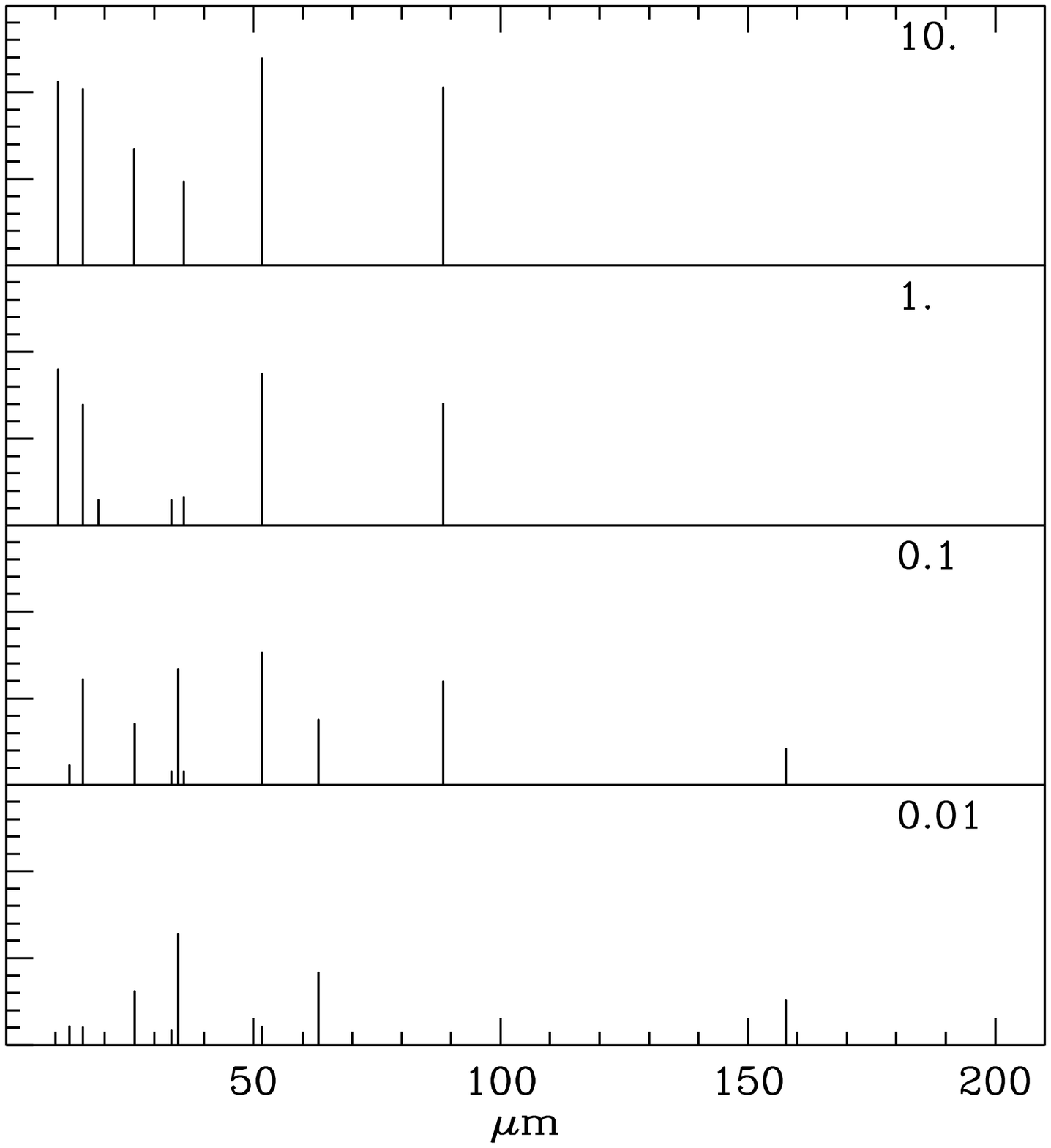}

\end{center}
\caption{Model calculations of line fluxes adopting a BB flux (Contini \& Viegas 2001b); 
see text}
\end{figure*}

The UV-optical lines appear in the
left diagrams of Fig. 2, while the IR-FIR  spectra are  shown in  the right diagrams.
The line absolute fluxes  calculated at the nebula  cover a relatively  large 
range  in order to distinguish between   strong and weak lines.
Thg line intensities are given in  log $\nu$F$_{\nu}$ (\erg) spanning four orders on the Y-axis 
scale of the right diagram  for each position. The  spectra in Fig. 2 which refer to the observed positions  
(Fig. 1, top diagram)  are  labelled on the left of the  diagrams.
The spectra  are constrained by the comparison of calculated (Contini 2009)  with 
observed (Simpson et al 2007) line ratios in the mid-IR.

It can be noticed  (Fig. 2, left diagram) in  the regions near the GC that  the \Ly and CIII] 1909 lines
are the strongest in the UV, while in the optical range [OII] 3727 and \Ha dominate
near the Arches Cluster. The [NII] 6583+6548 and [SII] 6717+6731 lines can also be strong in some
positions depending on the N/H and S/H relative abundances. In the IR range
[SIII] 0.94, [NeII] 12.8, and [SIII] 18.7 can be strong. [SIII]33.5 is generally the strongest line.
The [CII] 156 line  appears throughout most of the observed region.
Lines are  stronger in  regions corresponding to  higher densities (Paper I, fig. 2).

In  Figs. 3 and 4 we present a small selection of the spectra previously calculated
for AGN, starbursts and HII regions (Contini \& Viegas 2001a,b).
 The UV-optical lines appear in the
left diagrams as in Fig. 2, while the IR spectrum is  shown in the right diagrams.
The line absolute flux  scales are different in Figs.  3 and 4  in order  to display
even weak  significative lines (e.g. [CII] 156 and [NII] 203)  as well as the  entire \Ly line.

In Fig. 3  the models are suitable to  a narrow line region (NLR) of AGN, LINERs and LLAGNs.
The  spectra were calculated  adopting a power law (pl)  radiation  (Contini \& Viegas 2001a).
We have chosen the  parameter ranges close to those of the physical conditions   near the GC,
 as those calculated in Paper I
(e.g. preshock velocity \Vs=100 \kms and preshock density \n0=100 \cm3)
and  corresponding to different  intensities of the flux (indicated on top of the right diagram spectra)
 from the active center and different geometrical thickness of the emitting nebula $D$. 
The  top and  and bottom spectra
refer to  the matter bound and  to the radiation bound cases, respectively.
In particular,
 line absolute fluxes in \erg are calculated
 adopting a power-law radiation
with $\alpha_{UV}$=-1.5,  $\alpha_X$=-0.4  and different F$_{\nu}$.
\Vs=100 \kms, \n0=100 \cm3,\B0=10$^{-4}$ gauss and D=10$^{17}$ cm
(top diagrams), D=10$^{19}$ cm (bottom diagrams).
It can be noticed  that $D$ affects  the low-ionization level and neutral line intensity.

In Fig. 4  the spectra are calculated 
 adopting a black body (BB) flux  for  different ionization parameters $U$ 
(indicated on top of the right diagram spectra). The models are suitable
to  starbursts and  HII regions (Contini \& Viegas 1991b).
In particular,
 line absolute fluxes  in  \erg are
 calculated   adopting a black body radiation,
different $U$,  \Vs=100 \kms,
\n0=100 \cm3, \B0=10$^{-4}$ gauss, D=10$^{19}$ cm, and
\Ts = 5 10$^4$ K (bottom diagrams), \Ts=10$^4$ K (top diagrams).
%\Ts = 5 10$^4$ K (middle diagrams) and \Ts=10$^5$ K (top diagrams).
%\Ts = 5 10$^4$ K (middle diagrams) and \Ts=10$^5$ K (top diagrams).

The top and bottom diagrams correspond to different colour temperatures of the stars
which affect intermediate- and high-ionization level lines.

The \Ly line is the strongest in  Fig. 3 and 4 spectra. 
   Moreover,
the fine structure lines   emitted by several species in various ionization stages in the FIR
([CII]156, [OI]63, 145, [NII]122, [OIII]88, 51, [NIII]57) are the strongest lines in the IR spectrum
of most  galaxies  (Maiolino et al.  2009).

\subsection{Mid-IR lines}

We start  with  the analysis of the mid-IR lines, because  they were observed.

It was argued in Paper I that the spectra near the GC show relatively  strong
 low- and intermediate- ionization level
lines which  do not exclude a power-law dominated  flux from a weak central black hole.
Therefore, the lines   are compared with those emitted from  LINERs and low ionization AGN.
In  Fig. 5  we compare the  Spitzer  observations
of [OIV]26 \mum /[NeII]12.8 \mum versus [NeIII]15.5 \mum /[NeII]12.8 \mum
collected for a large sample of LINERs by Dudik et al. (2009) with the data observed by 
Simpson et al (2007)  near the GC.  
We will use in the present  figure and in the following ones 
 the numbered positions observed across the GC region (Fig. 1 top),  with an 
emphasis on those positions closest to the two major stellar clusters. 
Therefore, the data referring to the GC cover a  rather extended area of the diagrams
depending on the line ratios considered.

In Fig. 5  the Galactic  line ratios to [NeII] appear towards the lower tail of the 
  distribution,  indicating  a lower ionization level  of the emitting gas. 
However, these ratios involve  [OIV]/[NeII]. 
%To investigate 
%whether the slope and relative scattering of the data 
%depend on the relative abundance  O/Ne, we  show  in Fig. 5 (top diagram)    
% [NeV]14.3 /[NeII]12.8 versus [NeIII]15.5 /[NeII]12.8 observed by Dudik et al (2009).
%  The [NeV] line is too weak to be observed near the  GC so the data do not appear .
%Notice that  the distribution of the
%data in the two diagrams is  similar, indicating that  it  does not depend
%significantly on  O/Ne, but   mostly on  the physical parameters.
In particular, the intensity of the [OIV] line  is strongly affected by \Vs, as  found in Paper I.

Model  calculations  show  that  the [NeV]/[NeII] ratios are  all $<$ 10$^{-6}$ because
\Vs and $U$, which are the parameters responsible of  heating  and ionizing  the gas,
are relatively low.  Therefore, the [NeV] lines are omitted in Table 3. 

The general models (black asterisks connected by a solid line for a BB flux and
black crosses connected by a  a solid line for a pl flux) are also displayed on  Fig. 5. 
Comparison with starburst galaxies (BB models, Contini \& Viegas 2001b) 
show that the [NeV] lines 
are relatively high for \Ts=10$^5$ K, U $\geq$ 1 and \Vs $\geq$ 100 \kms. In shock dominated nebulae,
the [NeV] lines are strong for \Vs $\geq$ 200 \kms, and in AGNs (corresponding to pl dominated models,
Contini \& Viegas 2001a) they appear even for \Vs=100 \kms and
$F_h$ $\geq$ 10$^{10}$ cm$^{-2}$ s$^{-1}$ eV$^{-1}$.
Nevertheless.  the input parameters  would 
correspond  to higher \Vs ($>$ 100 \kms) and higher\ n0 ($>$ 100\cm3). 

Summarizing, Fig. 5 suggests that  the mid-IR line ratios near the GC are different from those of
LINERs and starbursts.
Comparison with models shows that the LINER  data are located between the BB and pl dominated models
indicating an hybrid character. 
 The modelling  shows that the GC data  are dominated by BB flux.

\subsection{Far-IR lines}

Far-IR (FIR) lines deserve consideration, even if   
  there were no possible FIR observations at the positions of Simpson et al,
because the intensity of fine structure lines is proportional to the heating rate, 
which is given by the UV radiation field
of young stars, in star-forming galaxies. The cumulative intensity of the strongest FIR  lines
is proportional to the star forming rate (Kaufman et al 1999).
 Walter et al (2009) show that the [CII] line is a critical tracer of star
formation in the first galaxies.
The relative intensity of FIR lines gives information on  metallicity
of the ISM (Nagao et al),  the density of the UV radiation field (Kaufman et al 1999),  on the
hardness of such field, and  AGN versus starburst excitation (Spinoglio et al 2005).

The [CII] 156 line  can provide a powerful cosmological tool to detect and characterize high-z galaxies.
Capak et al (2009) claim that
 a very promising means of confirming object redshifts
is through  [CII]
 because [CII] 156 line is the dominant gas cooling line in the interstellar medium,
traces the cold neutral medium and photon-dominated regions at the interface  with active
star forming regions, and  is now routinely detected at z$>$ 4  with existing sub-mm telescopes
(Maiolino et al. 2009, 2005).

Luninous FIR galaxies typically have   high dust-to-gas ratios. To
distinguish between silicate or carbon grains  the silicon features at
10 and 18 \mum  are observed. If the  carbon grain hypothesis prevails,
this could explain  unusually weak  [CII] lines indicating that carbon is locked  into dust grains.
However, the high-z galaxies  have enhanced [CII] emission by nearly an
order of magnitude. 
At high redshift [CII] is detected  in the sub-mm,  in the host galaxies of QSR
with L(IR) $>10^{12}$ (ULIRGS). 
High-z star-forming galaxies are observed to have lower gas-metallicity than local galaxies
(Maiolino et al 2008). 
In the  narrow line region (NLR) of AGN the metallicities   are solar or sub-solar, 
lower than in local massive galaxies.
Perhaps dust and carbon gas come from different regions
as suggested by Rubin (2009) who claims that 
the  effect of lower metallicity which leads to higher [CII] in low metallicity galaxies
is linked with reduced dust obscuration.

Model results for the GC (Paper I, fig. 2) show that   [CII]156  is relatively strong in the FIR.
[NII]203 is weak but present is the regions close to the Arches cluster.
The models correspond to HII regions, with star temperatures of $\sim$ 3 10$^4$ K and
moderate shock velocities of $\sim$ 70 \kms.
Calculated spectra for starburst and AGN (Figs. 3 and 4) show that the [CII] line 
intensity depends strongly on
the physical conditions of the emitting gas. It prevails at low shock velocities
in radiation dominated nebulae (large geometrical thickness) and 
 for relatively low photoionization flux intensities in both the
 BB and power-law models.

\subsection{Optical lines}

We will discuss the spectra from  regions near the GC in the UV and optical range
by the line ratios  presented in Tables 1 and 2, respectively.
They were calculated  using the  models which explain  the IR data.
These calculated spectra  are the most plausible substitute to the observational data.

  Comparing  the significant optical line ratios with those observed in HII regions,
Seyfert galaxies, LINERs, etc, (Veilleux 1985, Veilleux \& Osterbrock 1987, Ho et al. 1997)
we  obtain a further hint on the  nature of the spectra observed near the  GC.

The HeII 4686  and  HeII 1640 lines  are too  weak to be observed  near the GC.
This is due to the relatively low ionization parameter $U$ and the relatively low temperature
of the stars  \Ts.
The spectrum is similar to that of  low luminous  AGN (e.g.  NGC 4579, Contini 2004).
In fact the forbidden  lines from the first and second ionization levels are strong :
 [OII] 3727, [OIII] 5007+ ([OII]3727/[OIII]5007 $>$ 1) [NII] 6548+, [SII]6717+ in the optical
range,  and [SIII] 9400 in the NIR are present.

In Fig. 6 (top) the [OIII] 5007/\Hb~ versus [NII]/\Ha diagram shows the results obtained for the GC
in the different positions (filled magenta  circles).
The diagram is divided by the red lines in the  zones characteristic
of  Seyfert (S), LINER (L), HII (H) regions following  Veilleux et al (1987, Fig. 6a).

 The GC data  are scattered over the three zones. The  data   cover 38
 different positions, each with a beam size of approximately 250 square arcsec, and therefore sample
regions observed throughout a slit
crossing a limited region near the GC. We compare them with  2" wide long slit spectra  across  the Seyfert 2
galaxy NGC 7130 (Contini et al 2002, Radovich et al. 1997)  at  position angles PA=90$^o$ and 160$^o$
and with the inner 6" long slit spectroscopy at PA=11$^o$ and a high resolution spectrum in a 2" X 4"
region centered on the nucleus at PA=90$^o$ (Shields \& Filippenko 1990).
 These data  cover the
entire galaxy  including the active center and starburst regions,  central  and
peripheral.

  The slope of the line ratios determines the ionization source, however, the areas covered by  
Spitzer observations for the GC and for NGC 7130 are different by many orders of magnitude.
At the distance of the GC ($\sim$ 8 Kpc) the LH IRS Spizter beam   size is $\sim$ 0.5  pc$^2$.
At the distance of NGC 7130 (which SIMBAD  gives as 68.58 Mpc),
the beam  size would  be $\sim$ 1000 pc$^2$, dwarfing all of the GC observations.
So we have  integrated  single  emission lines  throughout all the 38 positions adopting
 that the physical conditions  of the filaments included in  each of the observed areas are the same.
The  weights  are calculated by w= (0.5/$\pi (D(pc))^2$) {\it ff} in each position.
The values of the geometrical thickness of the emitting nebulae $D$,  are given in Paper 1 (table 2),
the filling factor  {\it ff} =1.
The results are indicated as a large black dot on both Figs. 6 and  7   (top diagram).
Notice that the ISM regions have a strong effect on the  [OI]/\Ha line ratios beacause the models
are matter bound.

The presence of a Seyfert nucleus and  energetic  starbursts are evident from the optical
and near-IR properties of the galaxy.
Therefore NGC 7130  is a remarkable prototype of a "composite" galaxy (Contini et al. 2002).
The trend of the  NGC 7130 data is roughly the same as that of the GC.
 However the data  showing the maximum [OIII]/\Hb observed in NGC7130 are not reproduced
by the GC data. This  confirms that  the pl  photoionizing  flux in the central region
of   the Seyfert 2 galaxy  is stronger  than that  predicted in the observed regions near the GC
(Paper I, table 2).
 On the other hand, the BB radiation flux
 leads to  [OIII]/\Hb versus [NII]/\Ha line ratios  similar to
those of the  nebulae close to  starbursts.
For a low ionization parameter $U$, the trend of the   GC data in Fig. 6 (top diagram)
recovers that of  the HII regions in the
external regions of NGC 7130, far from the center,
which  are photoionized by  a BB flux from the  peripheral starburst stars.

\begin{figure}
\includegraphics[width=0.5\textwidth]{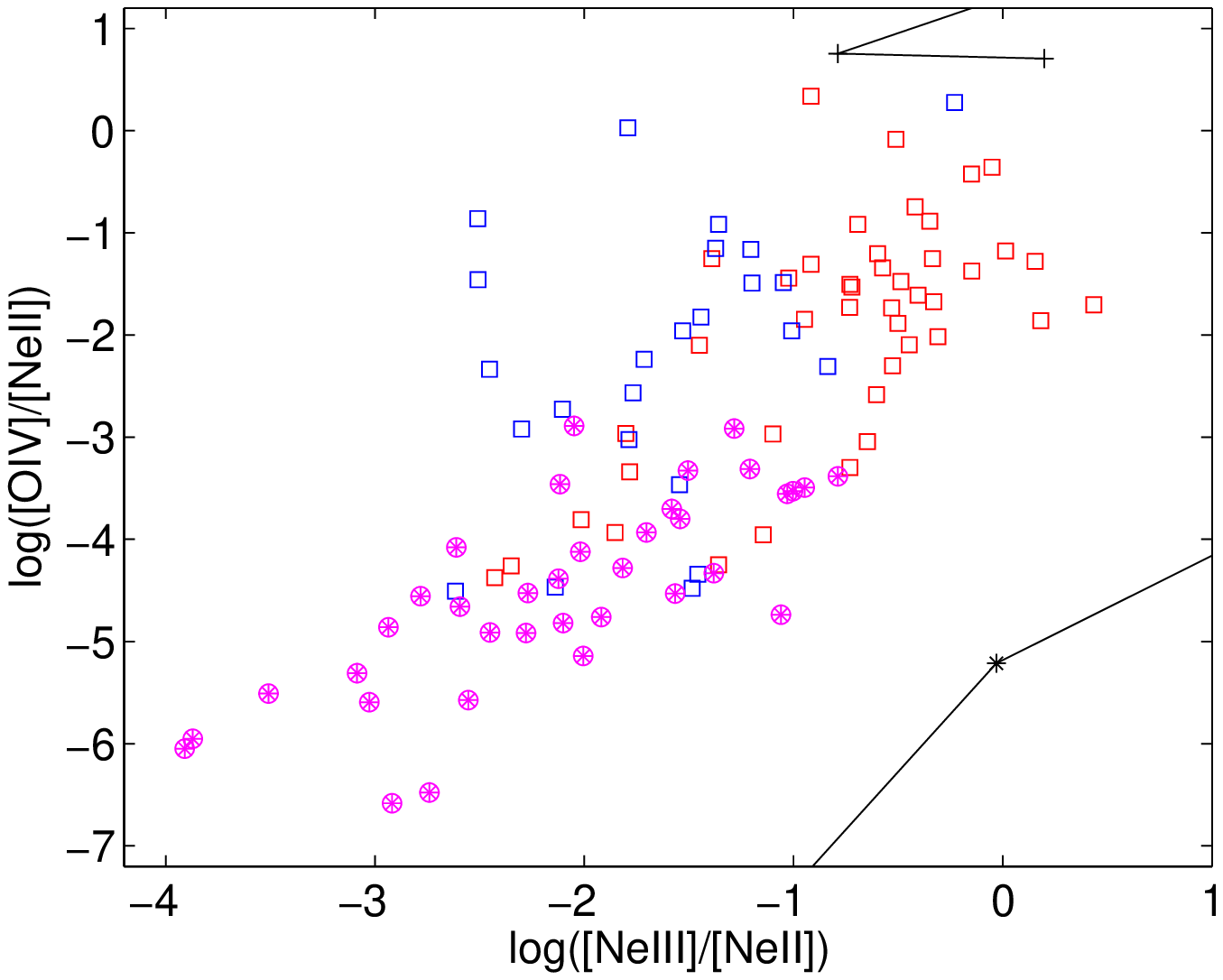}
\caption{Mid-IR spectra: red  squares and blue squares : IR-faint LINERs and IR-bright LINERs
  observations reported by  Dudik et al. (2009),
respectively.  Magenta filled circles : the GC observation by Simpson et al (2007).
}
\end{figure}

\begin{figure}
\begin{center}
\includegraphics[width=0.50\textwidth]{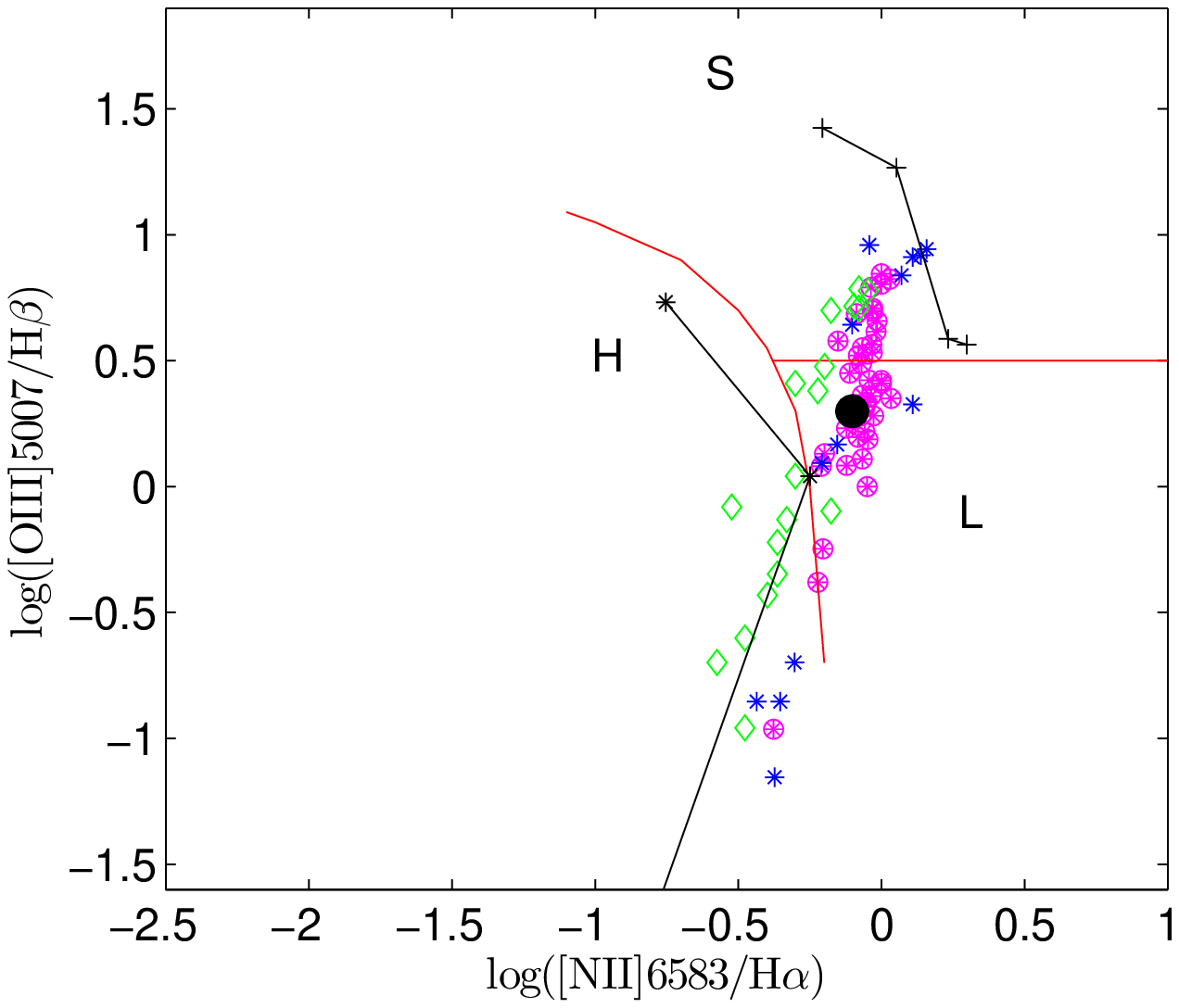}
\includegraphics[width=0.45\textwidth]{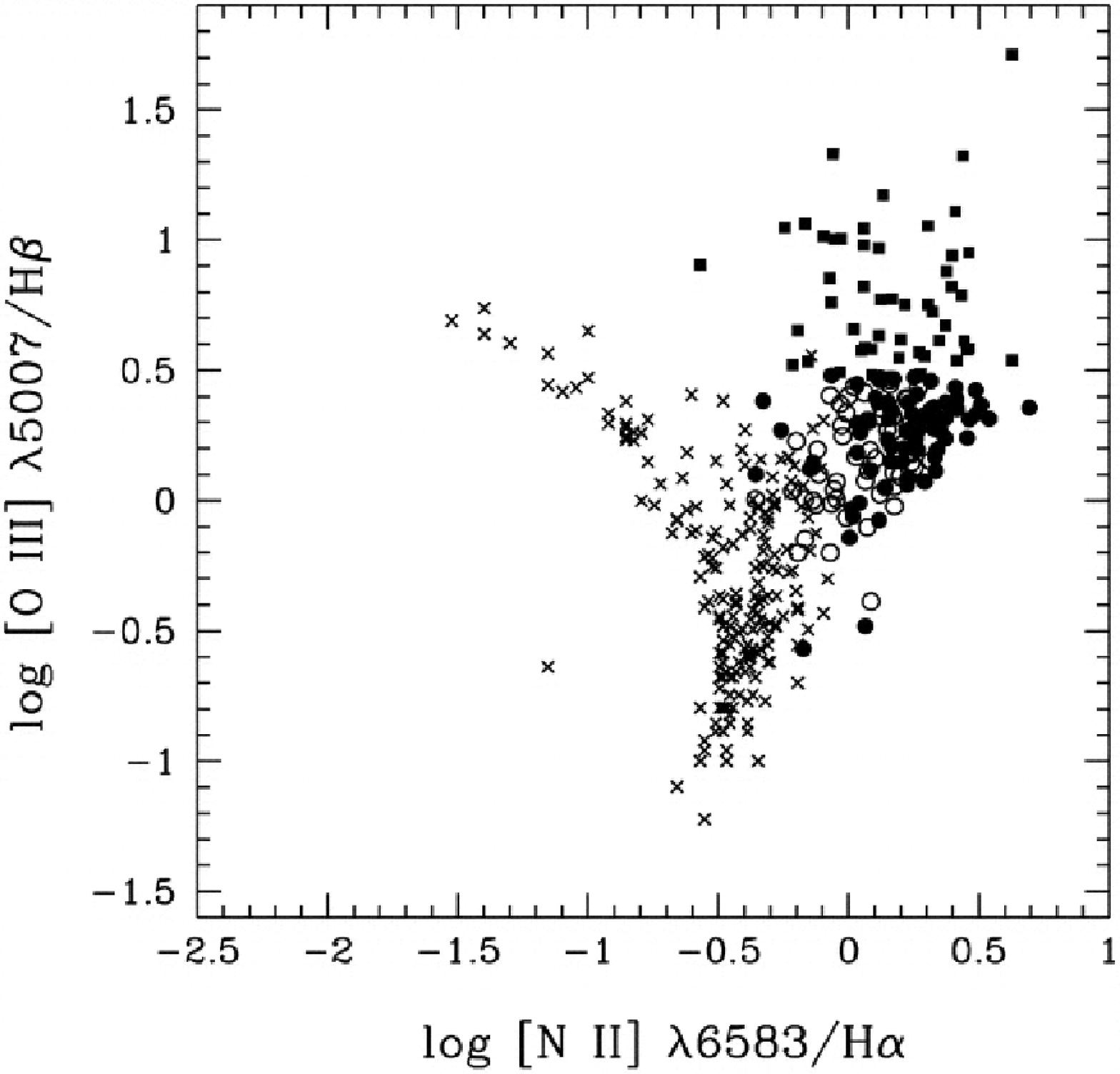}
%flushright
%includegraphics[width=0.30\textwidth]{veilleux.eps}
\caption{Top : comparison  with observed line ratios.
Magenta filled  circles :  model results for the Galaxy;
; blue stars : NGC 7130 (Radovich et al 1997); green open diamonds (Shield \& Filippenko 1990).
Bottom : [OIII]/\Hb versus [NII]/\Ha adapted from Ho et al (1997, fig. 7a) ;
  H II nuclei (crosses), Seyfert nuclei (squares), LINERs (filled circles), 
and transition objects (open circles). 
}
\end{center}
\end{figure}

\begin{figure}
%\begin{center}
%\includegraphics[width=0.50\textwidth]{milkopt31_n.eps}
\includegraphics[width=0.50\textwidth]{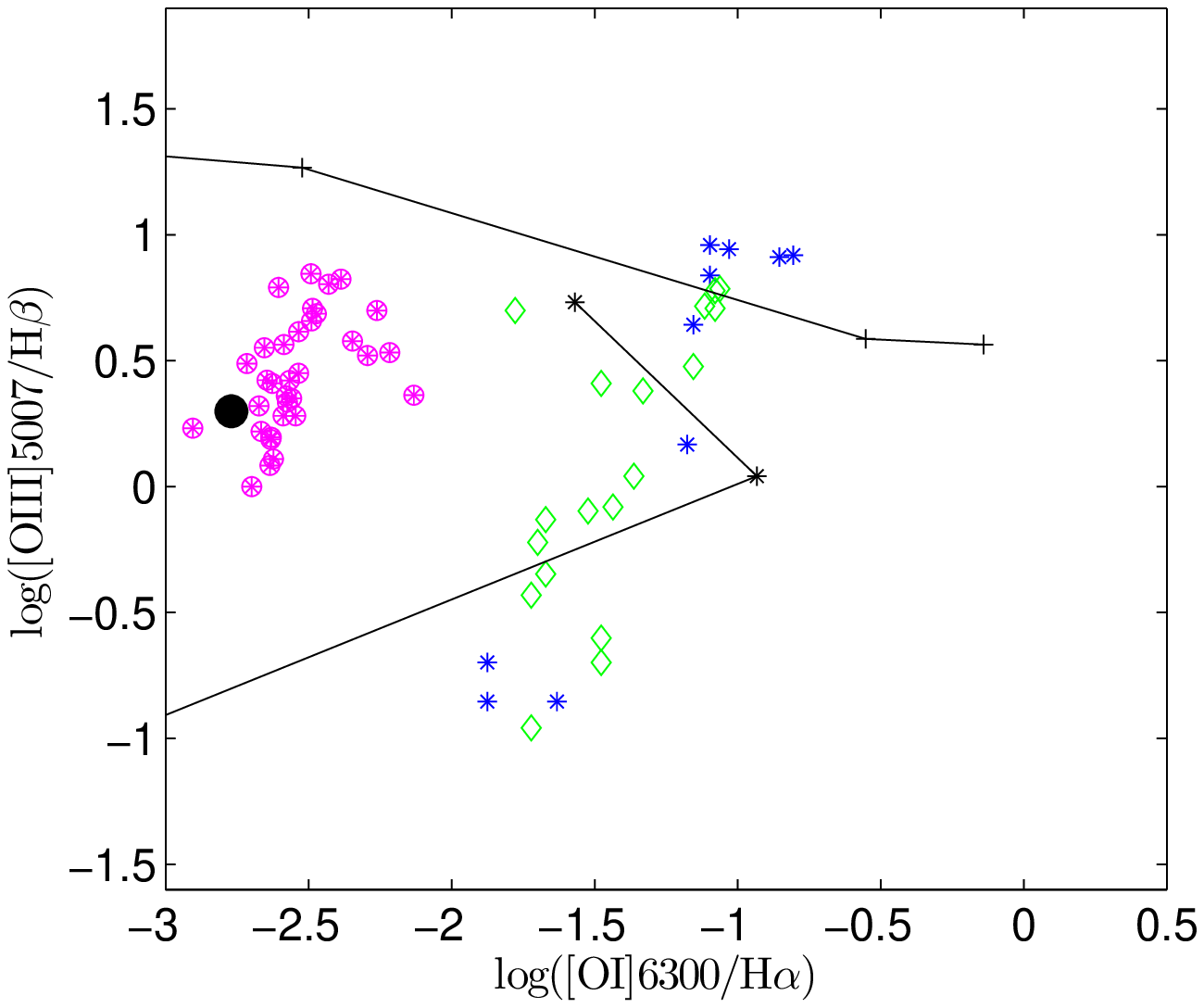}
\flushright
\includegraphics[width=0.48\textwidth]{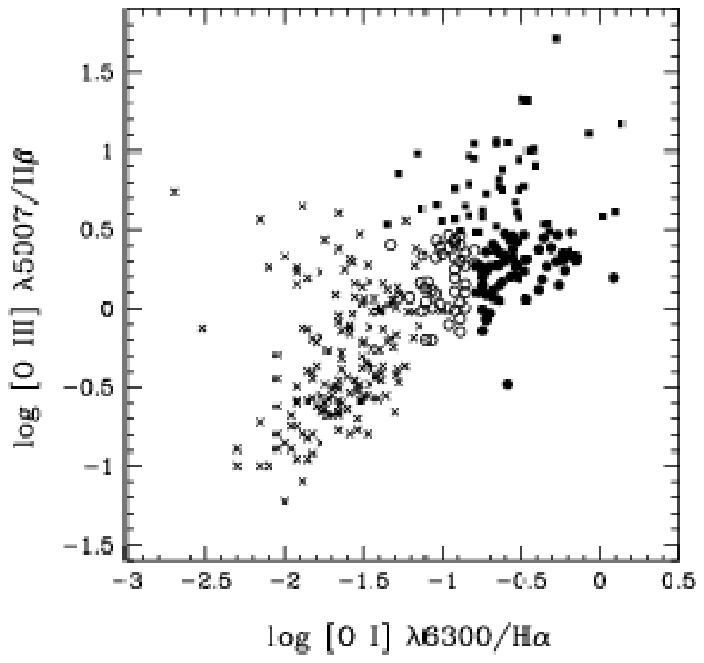}
\caption{Top : comparison  with observed line ratios.
Magenta filled  circles :  model results for the Galaxy;
; blue stars : NGC 7130 (Radovich et al 1997), data taken  at PA = 160$^o$ and 90$^o$;
open green diamonds (Shield \& Filippenko 1990), data taken at PA. = 11$^o$.5.
Bottom : [OIII]/\Hb versus [OI]/\Ha adapted from  Ho et al (1997, fig.7 c) :
symbols same as in  Fig. 6 (bottom).
}
%\end{center}
\end{figure}

The  models (black lines)
displayed on Fig. 6 (top  diagram)
 confirm that the GC data  imply $U$ $<$ 0.01.
 O/H and N/H relative abundances are about solar  for the GC spectra
 because the  models   are calculated with solar abundances.

In the bottom diagram of Fig. 6 we show for comparison
 the diagnostic diagram for a sample of active galaxies assembled by
 Ho et al (1997, fig. 7a).  The GC spectra  would appear in the regions
 suited to   HII (starburst galaxies). 

  In the bottom diagrams of Figs. 6, 7, and 8 we compare
  regions with very different  extensions (notice  for instance the scales in Fig. 1 top
and bottom diagrams)
and nature. The observations reported by Ho et al were taken with  large amplitudes  covering
the  entire galaxy. However, the   dominant character common to a certain group of objects (AGN, starburst, LINER, etc)
was confirmed by theoretical models. Therefore, we  consider that these comparisons are sensible.

To constrain the results, we compare  in Fig. 7 the [OIII]/\Hb versus [OI] 6300/\Ha (top diagram) 
calculated for the GC with the data observed throughout NGC 7130.
In the bottom diagram of Fig. 7 the  diagnostic diagram of  Ho et al (1997, fig 7c) is shown for comparison.
The GC data  show low [OI]/\Ha. This  confirms that the nebulae are filaments
with relatively small geometrical thickness ($D$), as calculated in Paper I,
 The fitting models are matter bound because the cooling rate in  the  post-shock region is low.
The downstream  edge  of the nebula is reached  before the
 gas cools to temperatures ($<$ 10$^4$ K) low enough to recombine  emitting  the bulk of  neutral lines.

 This is an important issue  revealing that matter is highly fragmented.
It  could be a result of   Richtmyer-Meshkov  (RM) ( Mikaelian
1990; Graham \& Zhang 2000)
 instability which is an interfacial instability between two fluids of different densities driven
by weak and intermediate shock waves
characteristically leading to photoionized  "fingers". (For a discussion of the 
fragmentation generated by Kelvin-Helmholtz instablities, see Sect. 3). 
The "ubiquitous ISM filamentary structure" is
confirmed by Molinari et al. (2010) by the Herschel Hi-GAL Milky Way survey.
This suggests fragmentation which
derives from a turbulent regime.

\subsection{UV lines}

\begin{figure}
\begin{center}
\includegraphics[width=0.45\textwidth]{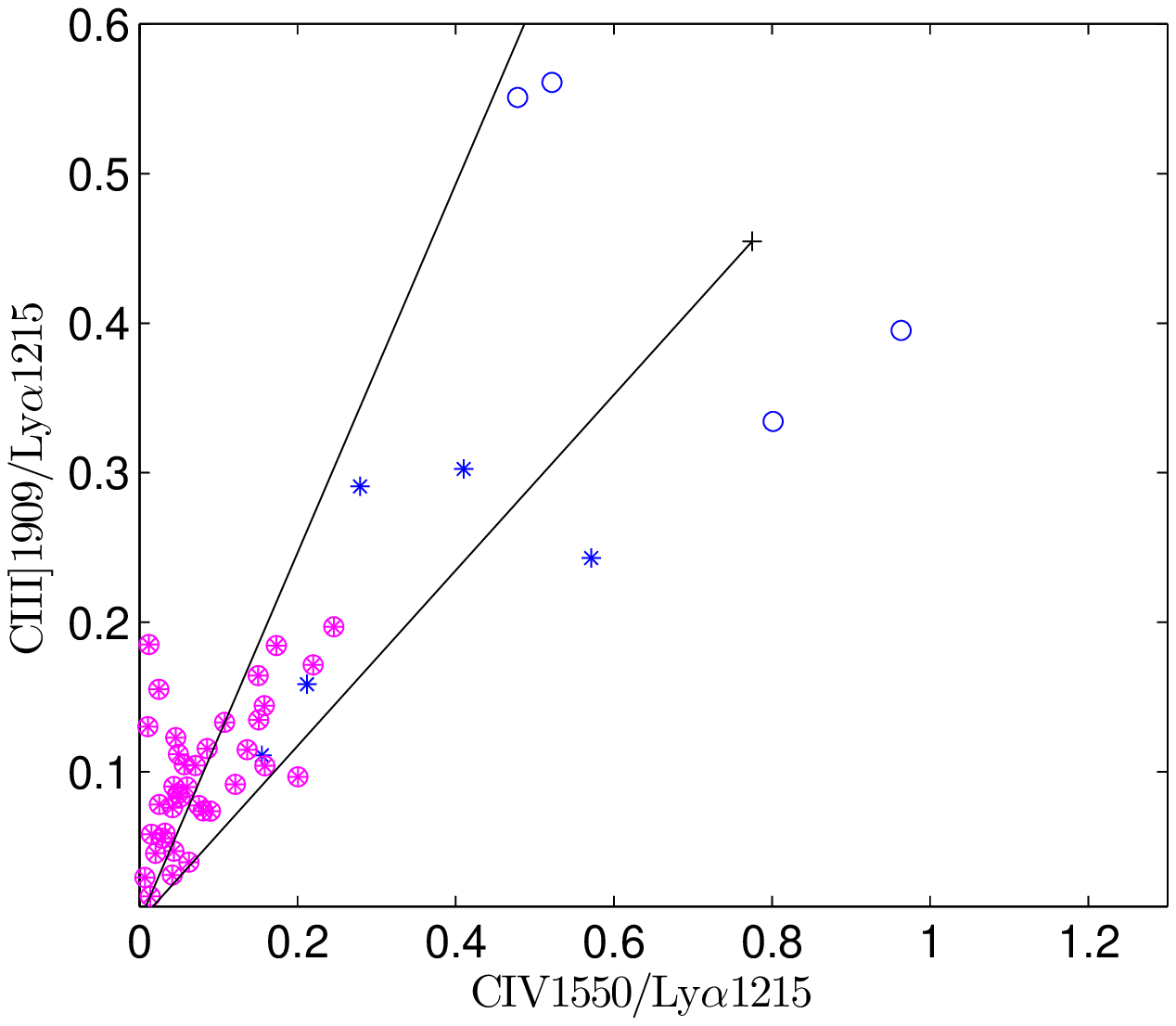}
\includegraphics[width=0.38\textwidth]{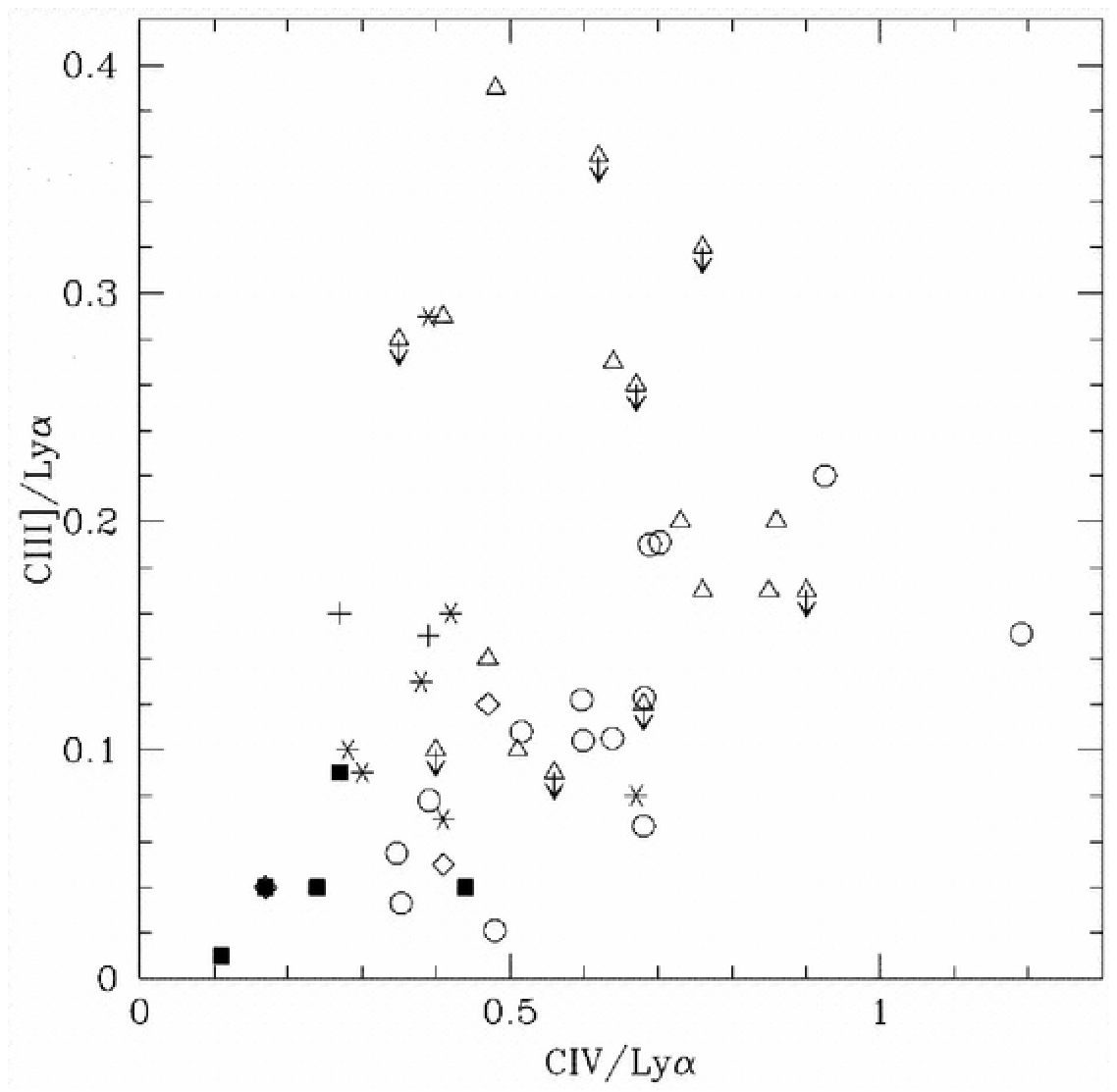}
\caption{Top:  magenta filled circles  represent the GC data
(Table 1);   
Contini \& Viegas (2001a,b) BB models are represented by black solid lines  
and pl models by black solid lines connecting black crosses;
blue open circles and asterisks from Kraemer et al (2000), SW and NE data, respectively.
Bottom :
UV spectra from  NLS1 (black squares), Seyfert 1 and  quasars from
Kuraszkiewicz et al (2000, fig. 1)}
\end{center}
\end{figure}

The  evolution of  star forming galaxies  and quasars
has been  investigated in the last years (e.g. Atek et al 2008 and references therein)
through numerical simulations, focusing  on the \Ly line 
  which is considered the strongest observable line from high-redshift galaxies (Laursen 2009).
The calculated spectra  for  AGN and starburst galaxies  based on the models used in this paper
 are presented in Contini \& Viegas (2001a,b).
The low \Vs, low \n0 models appear  in graphical form in Figs. 3 and 4, respectively.
The left diagrams  in Figs. 3 and 4 show that \Ly  is  by far the strongest calculated UV line. 

The models show that, although  \Ly line always prevails,  its ratio to the UV lines
(e.g. CIII]1909) is lower for HII regions  than for AGNs.
However,  recall that the  spectra are calculated using cosmic abundances, so
line ratios referring to  different elements can be different, depending on the relative abundances.

Bertone et al (2009) claim that in the low redshift universe the \Ly forest
may trace less than half the baryons. They suggest that the gravitational shocks, produced 
at the time of  structure formation,
progressively heat  the diffuse intergalactic medium.
Approximately half of the intergalactic gas is predicted to have temperatures in the range of
10$^5$-10$^7$ K at z $\sim$ 0 which is too hot to emit or absorb hydrogen \Ly radiation.
However, this shock heated gas (hot-warm IG medium) has not been unambiguously detected.
Metal lines in the UV band could detect the cooler fraction of the hot-warm gas.
The models presented by Contini \& Viegas (2001a,b) accounting for the shocks
give a hint on the ratio of UV lines to  \Ly.

The CIV 1550 and \Ly line are  the strongest in the UV from gas heated to 
10$^4$-10$^5$ K by primary and secondary
radiation  due to photoionization and emission from slabs heated by the shocks, respectively.
This can be noticed in different objects such as AGN, starbursts, etc and, on a small scale,
supernova remnants, novae, and symbiotic star systems etc.
In the GC the  CIII] 1909 multiplet is also important because photoionization is relatively low.
                 
In Fig. 8  CIV/\Ly  versus CIII]/\Ly is shown for the GC (top) and compared with   NLS1 
(Narrow Line Seyfert 1) and other type of galaxies (bottom) in the diagram presented by
Kuraszkiewicz et al (2000, fig. 1).
We have added in the top diagram the data in  regions observed by  Kraemer et al (2000)
at different  locations throughout  the Seyfert 2 galaxy NGC 4151 (Contini et al 2002).
 The observed spectra were obtained by low-dispersion long-slit data, at a position angle P.A. = 221°, using 
the Hubble Space Telescope Space Telescope Imaging Spectrograph,
using  a narrow slit  width of 0".1, and extracting fluxes in bins with lengths of 0".2 in the inner 1"  
and 0".4 farther out.

The line ratios for the GC are  different and separated from those of the AGN.
They roughly follow the trend of the BB dominated models.

The UV  spectra calculated for the  the GC regions  show many lines with  fluxes 
of $\sim$ 10$^{-4}$ and 10$^{-3}$ \erg for 
\n0 $\sim$ 3 and \n0 $\geq$ 100 \cm3, respectively  (Fig. 2). 
The star temperature \Ts is $\sim$ 3 10$^4$ K and $U$ $\sim$ 0.01 (Paper I, Table 2).
The  HII region models (Contini \& Viegas 2001b) with \Vs=100 \kms and \n0=100 \cm3 adopted to reproduce the   
UV line ratios  correspond to lines  with absolute fluxes between 10$^{-1}$
and 1 \erg mostly  for \Ts=10$^5$ K and U$>$0.1 (Fig. 8, top). 

\section{Turbulence, abundance fluctuations, and magnetic field} 

The analysis  presented in Paper I invoked the existence of shocks as a necessary ingredient
for a consistent derivation of the spectra observed by  Simpson et al (2007).
Shocks are known to produce turbulence, notably via the Richtmyer-Meshkov instability  
which is due to the shock acceleration, and 
  is similar to the familiar Raleigh-Taylor instability (see e. g.  Mikaelian
1990; Graham \& Zhang 2000 ). Shocks can also lead to other instabilities e.g. 
shear instability, and Kelvin Helmholtz instability which in turn generate turbulence.
Indeed, the clumped and fragmented morphology of the region - that was confirmed in Sect. 2.3 explaining the
[OI]6300/\Ha line ratios -   strongly  suggests   
the existence of an underlying   supersonic shock-generated turbulence. 

 Interestingly, a turbulent regime was already  predicted by Simpson et al  from
the characteristic morphological structure of the complex region near the GC.

\subsection{Methodology}
 
The existence of turbulence can be probed  directly  by analyzing the velocity spectrum. In addition, the imprint of turbulence may be detected   indirectly  by analyzing the power spectrum of 
"passive scalars" (Lesieurs 1997), which are 
strongly coupled to the gas and follow its turbulent motion   (but do  not feed-back on the 
 turbulence). The power spectrum of the passive scalar is
proportional to the turbulent velocity power spectrum. 

Examples are the  21 cm emissivity which is proportional the column density of the neutral hydrogen and its fluctuations reflect those of the density fluctuations and in turn the velocity fluctuations . In the case of infrared continuum it is the dust density, and in the case of abundances ratios it is the density of the specie under consideration. 

The observational  power  spectrum of a passive scalar is a way to measure  its
  hierarchical spatial structure.  In cases that it is a power law   a hydro-turbulence  is naturally suggested as the
mechanism that has generated the observed spatial structure. Nevertheless, it is not enough to prove the existence of a  hydrodynamical turbulence. On the other hand, if the power spectrum of the velocity field reveals a turbulence, the fact that 
 the power spectra   of passive scalars are also power laws with the same slope strengthens the 
 credibility of the deduced velocity turbulence.
  
   An example is the power spectrum of 
21cm emission in the small Magellanic cloud (Stanimirovic et al 1999). It was shown by Goldman (2000) that the power spectrum is consistent 
with that generated by a large scale velocity turbulence. 
Additional support for the existence of a dynamic turbulence was obtained by Goldman (2007) by deriving the power spectrum of 
the radial velocities of the giant H$_ I $ super-shells of the SMC.

In what follows we analyze the radial  velocity data of Simpson et al (2007), and their observed 
mid-IR continuum flux in order to test for the existence of such a turbulence. 
Indeed  these two kinds of observational data reveal the existence of a supersonic turbulence.

Following these two analyses, we examine the  Si/H abundance as function of position  
{\it  computed} in Paper I. The idea being that if these computations indeed reconstruct   
the actual  physical conditions in the region then
the imprint of the turbulence should be evident in the computed abundance as function of position. 
It turns out to be indeed the case, thus lending 
  credibility to
the computations of Contini (2009) as well as to their extensions  to  the optical and UV ranges 
presented in previous sections of the present paper. 

Finally, we estimate the effect of turbulence on the magnetic field.

\subsection{Radial velocities}

The  3D spectral function of the turbulent velocity, $\Phi(\vec{k})$ is defined in terms of the 2-point
autocorrelation of the turbulent velocity field

$$ 
\Phi(\vec{k}) = \frac{1}{(2 \pi)^{3/2}}\int <\vec{v}(\vec{r'}) \cdot \vec{v}(\vec{r}+\vec{r'})>   e^{i \vec{k}\cdot \vec{r}}  d^3r$$
In  the homogeneous and isotropic case it is useful to introduce the
turbulence energy spectrum $E(k)$ and the turbulent velocity spectral function $F(k) = 2 E(k)$ so that

$$\Phi(\vec{k})=\Phi(k)=\frac{  F(k)}{4\pi k^2}\ ;  k=| \vec k|$$
Assuming the ergodic principle, ensemble averages can be replaced by space, surface, or as in the present case line averages. 

$$c_r(x)=<v_r(x')v_r(x'+x)>= \frac{1}{L}\int_0^L  v_r(x')v_r(x'+x)dx'$$

Here, the observed velocity $v_r(x)$ is in effect an intensity-weighted average
of the velocity along the line of sight. Under the assumption of homogeneity along the line of sight, the observed velocity is proportional to the integral of the velocity along the line of sight.
The resulting value is contributed by all the geometrical depth along the line of sight in the optically thin case  or  the optically thin part  of it.

 $F_r(k)$,  the power spectrum (the spectral function) of the radial turbulent velocity 
  is the one dimensional Fourier transform of $c_r(x)$ 
 $$F_r(k)= \frac{1}{(2\pi)^{1/2}}\int c_r(x)e^{ ik x} dx$$

The power spectrum can also be evaluted by,
$$F_r(k)= |v_r(k)|^2$$
where $v_r(k)$ is the Fourier transform of the velocity
$$v_r(k)= \frac{1}{(2 \pi)^{1/2}}\int  v_r(x) e^{ ik x}dx$$

Since the data  are given at a set of discrete positions 
we compute the power spectrum by the sqaured absolute value of the discrete Fourier transform of the velocities . This yields the a discrete power spectrum 
as function of the discrete wavenumber defined as $ 2\pi/l$, with $l$ being the corresponding spatial scale.

  The data are given for 38 positions along an almost straight line
with extension of about 75~pc. Not all positions are evenly spaced , thus a numerical
  uncertainty in the small-scales part of the power spectrum is expected.
  
Also, the radial velocity observed by Simpson et al (2007) for the m20 position  is substantially larger than the
adjacent velocities and most likely  is due  to a local outflow,  not typical of 
the general velocity field. We therefore adopt for this point a value equal to the mean velocity of the other positions.

In Fig. 9 we show the power spectrum of the radial velocities fluctuations, with respect to their mean value.
For random residuals the expected power spectrum would have shown no dependence
on the wave-number.
In contrast,
the power spectrum shown in Fig. 9 is a rather steep decreasing function of the relative wavenumber $k= \frac{l_0}{l}$. Here $l$ is the scale corresponding to $k$, and  $l_0$ is the largest scale   corresponds to $k=1$. 
 The largest scale  equals the linear extent of the data strip: about $l_0=75$~pc.
The value of the root mean square turbulent velocity on the largest scale is $v_0 = 17.4$ \kms and exceeds the 
thermal velocity of $\sim 10$ \kms.
We believe  that the scatter in the points of the power spectrum for the small scales ($k\geq10; l \leq 7.5 pc$) is an artifact of the uneven spacing, mentioned above.

The line in the figure is a $k^{-2}$ power law that is expected for supersonic turbulence (Passot, Pouquet, 
\& Woodward, 1988; Girimaji \& Zhou, 1995). It differs from
$k^{-5/3}$ Kolmogorov spectrum which characterizes incompressible turbulence. The steeper slope
is due to the fact that a fraction of the turbulent kinetic energy density at a given wavenumber is converted to compression work and decreases the energy transfer to the larger wavenumbers.

Indeed the spectrum derived  here is consistent with the $k^{-2}$ spectrum of supersonic turbulence. 
 In order to obtain a quantitative estimate of the goodness of the fit we adopted  an uncertainty 
of $15 km s^{-1}$  in accord with Simpson et al (2007). Then we generated synthetic velocity data 
by random generating at each position    a velocity using a normal distribution with a mean equal 
to the observed values and standard deviation of  15 \kms. 

We have built 100 such sets. For each we computed the power spectrum and then computed the standard deviations of the 
logarithm of the ensuing 100 power spectra. A log-log rather than a linear computation was done as a linear one would completely under-represent the high wave numbers (small spatial scales).

Doing so we obtain a value of $\chi^2 =0.72$. Some of the uncertainty in the power spectrum is due to 
the uneven spacing of the positions. The fact that even so the fit is a good one can imply that 
the true uncertainty in the observed velocities is {\it smaller} than  15 \kms.

If one assumes further that the turbulence is isotropic, the value of rms turbulent velocity  
corresponding to the largest scale would be $ \sqrt{3} \times 17.4 =  30$ \kms.

The characteristic timescale of the turbulence is given by $\tau_0\sim l_0/v_0 \sim 4 \times 10^6\ yr$. 
Namely,   comparable to the age of the young stellar populations. 
This is consistent with the framework adopted by Simpson et al (2007) and by  Contini (2009) that 
stellar outflows are the main energy  source that shaped  the ISM in this region.

 \begin{figure} 
%  \centerline{\includegraphics* [scale=1]{turb1_err.eps}}
  \centerline{\includegraphics* [scale=1]{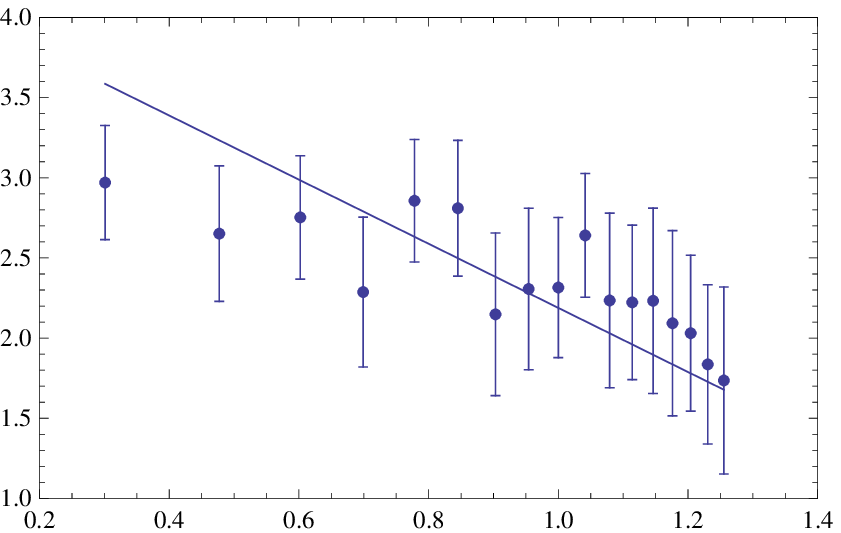}}
\caption{Log of the  power spectrum of the radial velocity residuals, in   units of (\kms)$^2$, as function of the   
log of the relative wavenumber k. $k=1$ corresponds to the spatial scale $l_0 = 75 pc$.  The line is a power law with index -2. }
% \end{figure}

%\begin{figure}
%\centerline{\includegraphics*[scale=1]{turb2.eps}}
\centerline{\includegraphics*[scale=1]{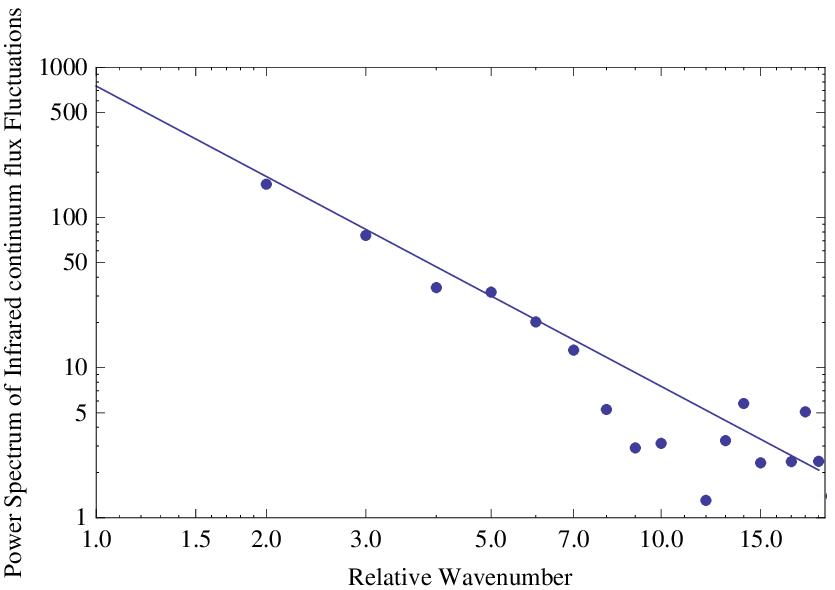}}
\caption{The power spectrum of the fluctuations of the mid-IR continuum at $(13.5-14.3)\mu m$, in   units of Jy$^2$, as function of the
relative wavenumber k. $k=1$ corresponds to the spatial scale $l_0= 75 pc$.  The line is a power law with index -2. }
% \end{figure}

% \begin{figure}
% \centerline{\includegraphics*[scale=1]{turb3.eps}}
 \centerline{\includegraphics*[scale=1]{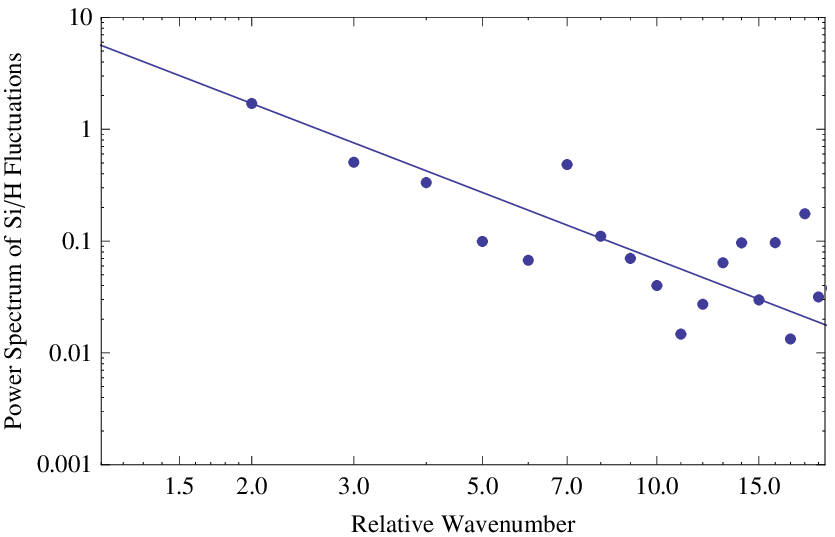}}
\caption{   The power spectrum of Si/H fluctuations, in  units of $10^{-10}$,  as function of the relative
wavenumber k. $k=1$ corresponds to the spatial scale $l_0= 75 pc$.  The line is a power law with index -2. }
 \end{figure}

Simpson et al (2007) estimate the uncertainty in each radial velocity determination to be $10-20$ \kms.
Thus, the deduced turbulent velocities from the power spectrum analysis are uncertain by 
values that are $38^{-1/2}$ smaller namely $\sim 1.6-3.2$ \kms.
For a relative wavelength of 16 ( scale length of about 2.5 pc)   the turbulent velocity is $8.7$ \kms larger than the uncertainty by a 
factor of $\sim 2.7-5.4$. Therefore, the derived power spectrum down to these scales can be trusted.

The power spectrum suggests a steepening of the slope to $k^{-3}$ at about $k=7$. If real, it implies that line of sight
depth is about $\frac{1}{7}$ of the extent on the plane of the sky (Goldman 2000); namely a depth of $\sim 10 \ pc$. 
This is in line with the conclusion of   Simpson et al (2007)  that 
there is a non negligible depth to the region. 

 \subsection{Mid-IR flux}

The observed flux is    an    integral of the flux along the line of sight.
It is contributed by all the gemetrical depth along the line of sight in the optically thin case  or  the optically thin part  of it.

 Miville-Desch{\^e}nes, et al. (2007, 2010) have demonstrated that the fluctuations of the far-IR continuum flux from thermally 
heated dust in the ISM exhibits a power spectrum identical to that of the velocity turbulence. 
The interpretation of this result was that the dust is coupled to the gas (Paper I) and thus the dust density 
fluctuations are determined by the gas turbulence. In the cases considered by these authors the turbulence was subsonic 
and the power spectrum was the Kolmogorov spectrum.
 
Here we wish to pursue a similar analysis. But since the turbulence is supersonic we anticipate that also the flux power  
spectrum would   also exhibit the $k^{-2}$ dependence. We use the  average   continuum flux   in the   range $(13.5-14.3)\mu m$, as 
function of position,  observed by Simpson et al (2007) and derive the power spectrum of the fluctuating flux. 
We do so by following the same procedure as for the radial velocity with the   flux being the quantity whose one-dimensional
autocorrelation is being Fourier transformed.

The resulting power spectrum is displayed in Fig.~10. The spectrum exhibits a clear $ k^{-2}$ behavior  for the larger scales 
and here too there is an indication of a steepening at about $ k=7$ to a $k^{-3}$ dependence; 
in effect it is clearer than in the radial velocity spectrum.  The uncertainties reported by Simpson et al (2007) are 
 quite small (of the order of $1\%$ ). As a result very small  uncertainties in the power spectrum 
 are obtained. In this case it is clear that the uncertainties in the power spectrum are dominated by the uneven spacing of the positions. 
 
It is possible to estimate the uncertainties of the uneven spacing by generating synthetic data with randomly spacings 
(much in the manner we did for the velocity uncertainties as reported above). 
However this seems to be outside the scope of the present paper. 
Therefore, we do not derive here a quantitative estimate for the goodness of fit. 
Rather we would like to draw attention to the fact that the deviations from the power law are at the smaller spatial 
scales, as expected from uneven spacing, and not on the largest scale. We use this power spectrum as means for 
lending support to the existence of the supersonic turbulence.

 \subsection{Si/H abundance}
 In Paper I the  various abundances are computed as function of position.
If these computations are a fair representation of the physics in the region, then one expects that the computed values will reflect the turbulence revealed in the observational radial velocity and mid-IR flux.

The abundances can be regarded  as a  turbulent  ``passive scalars'' (Lesieurs 1997).  Therefore, the positional  fluctuations in the 
observational  abundances can be due to the underlying turbulence.
To test this possibility, we chose one of the abundances (Si/H) and computed the power spectrum 
of the 2 point correlation of its fluctuations, around the mean. We selected Si/H  because it does not show drastic variations with position which (
as e.g. Fe/H), would imply trapping into dust grains and eventual sputtering.

The computed value is    an    integral of the abundance along the line of sight.
It is contributed by all the geometrical depth along the line of sight in the optically thin case  or  the 
optically thin part  of it.

The resulting power spectrum is shown in Fig. 11. 
The power spectrum is consistent with the $k^{-2}$ spectrum of the radial velocity turbulence and that 
exhibited by the mid-IR flux. This lends credibility to the computational model employed in  Paper I and
which is used also in the present work. 
Here, too there is an indication of a steepening of the spectrum at $ k \sim 7$ to a $k^{-3}$ dependence.\\
  In this case too, the deviations of the power spectrum from a power law are manifested mainly on the smaller scales. 
Here too, the consistency of the power spectrum with what is expected from an underlying supersonic turbulence strengthens 
the case for the existence of the latter. 
 \subsection{Turbulent Magnetic Field}

The high value of the 3-D turbulent velocity can amplify a preexisting magnetic fields up to equilibrium values  Shu (1992):
$$\frac{1}{ 8\pi}B_{turb,eq} = \frac{1}{2}\rho v_{turb}$$
where $\rho$ is the gas density and $v_{turb}$ is the turbulent velocity on the largest scale. For $v_{turb}$ =30 \kms

$$B_{turb,eq}=140\ \mu G \  (n/ 100cm^{-3})^{1/2}$$
 with 
 $n$ denoting the gas number density. The amplification 
proceeds via the dynamo mechanism involving the winding up of magnetic field flux lines by the turbulence eddies. 
Note that on the largest scale the amplification could have been a factor of few because the time available is of the order of turbulence timescale. On smaller scales, the turbulence timescale is shorter and the amplification is larger, but the equilibrium value is smaller.

Fields of this strength and even an order of magnitude higher were reported for the galactic 
center region (see review by Vall{\'e}e, 2004). The equilibrium value estimated here refers to the 
turbulent magnetic field. The larger reported values could evolve from the turbulently 
amplified field by an inverse cascade process (Mininni 2007) from small to large spatial scales, or directly by shock compression of the
magnetic flux lines embedded in the
ionized ISM (Medvedev, Silva, \& Kamionkowski  2007).

\section{Concluding remarks}

In a previous paper (Paper I) we have calculated the physical and 
chemical conditions in the nebulae heated and ionized by the star clusters
near the GC by   detailed modeling the mid-IR  line and the continuum spectra observed in 38
positions by Simpson et al. (2007). 

Summarizing,  in Paper I we found that
the spectra near the GC show  relatively low ionization due to a low average temperature
of the cluster stars (27000-39000 K). The gas pre-shock density and velocity are also low
($\sim$ 10-100 \cm3 and 70-150 \kms, respectively),  the geometrical thickness of the
emitting filaments is rather low ($\sim$ 0.01 pc), and the Fe/H relative abundance is inhomogeneous.
In certain positions, e.g. close to the Arched filaments, Fe is trapped into dust grains.
The initial magnetic field  ranges from $5\times 10^{-6}$ to $9\times 10^{-5}$ Gauss.

The UV and optical lines cannot be observed  due to  heavy extinction
at these wavelengths.
In this paper,   spectra were calculated in
the UV-optical-IR ranges, using  the physical  parameters and relative abundances
 which successfully reproduced the MIR data in Paper I

The line ratios calculated  for regions
 near the GC are compared  with those  observed  throughout single galaxies, in particular  Seyfert 2 galaxies.
We have chosen NGC 7130, which contains many starbursts close to the center and in
the peripheral zones, and NGC 4151,  which has been observed at different positions.
 The characteristic line ratios  in different ranges are also discussed
in the light of   the diagnostic diagrams  assembling large samples of active galaxies.

We have found  that the characteristics of the nebulae near the GC are different
from those of starburst galaxies and of the NLR of AGN.
As a  direct consequence of the low density in the nebulae near the GC 
the line intensities are low compared
to active galaxies. Low compression downstream of shock fronts
accompanying the  low velocity, low density nebulae, lead to the characteristic spectra
 (Fig. 2),  in which e.g. the \Ly line ratios to the strongest
UV lines  (CIV, CIII] etc) are relatively low compared to those of starburst and
AGN ratios (Figs. 3 and 4).

 The results show that the geometrical thickness of the emitting filaments is particularly 
small  revealing fragmentation of the emitting matter.
 The clumped and fragmented morphology of the GC region   strongly  suggests that it has been shaped
 by a supersonic turbulence generated by the shocks implied in the modelling (Paper I).
 Shocks are known to produce turbulence, notably via the Richtmyer-Meshkov instability  
which is due to the shock acceleration, and 
  is similar to the familiar Raleigh-Taylor instability (see e. g.  Mikaelian
1990; Graham \& Zhang 2000). Shocks can also lead to other instabilities e.g. 
shear instability, and Kelvin Helmholtz instability which in turn generate turbulence.
 
 The existence of such a turbulence was  confirmed  in Sect 3.  in the power spectra of the 
observational radial velocities and mid-IR continuum flux. 
The turbulence radial rms velocity is about 17 \kms and its 3D value is about 30 \kms.
 The associated turbulence timescale is about 4 Myr -   comparable to the ages  of the young 
stellar populations which are  ultimate generators of the shocks and the ensuing turbulence. 
The turbulence can amplify the initial magnetic fields (Contini 2009) by an order of magnitude.
 
 The power spectra exhibit a $k^{-2}$ behavior typical to supersonic turbulence, and  steeper 
than the Kolmogorov spectrum appropriate for incompressible turbulence.
The steepening of the power spectra for relative wave numbers exceeding 7, suggests that the line of 
sight depth of the turbulent region is about 10 pc.

Interestingly,  
  the  Si/H abundance as function of position,  {\it  computed} in Paper I, exhibits a power spectrum 
which is quite similar to the observational power spectra. This  lends credibility to  
computational model employed in  Paper I and in the present paper.

\section*{Acknowledgments}
 We are  grateful to  the referee for  many interesting comments which improved the 
presentation of the paper. We thank Sharon Sadeh for helpful advise.\\
IG thanks the support from the Afeka College Research Committee.

% \begin{thebibliography}{99}
\section*{References}

\def\ref{\par\noindent\hangindent 20pt}
\ref Allen, C.W. 1976 Astrophysical Quantities, London: Athlone (3rd edition)
\ref Atek, H.; Kunth, D.; Hayes, M.; ?–stlin, G.; Mas-Hesse, J. M.  A\&A  2008, 488, 491 
\ref Bertone, S., Schaye, J., Booth, C.M., Dalla Vecchia, C., Theuns, T., Wiersma, R.P.C.
2010, arXiv:1002.3393
\ref Capak, P. et al 2009, arXiv:0910.0444
\ref Contini, M. 2009 MNRAS, 399, 1175, Paper I	
\ref Contini, M.  2004 MNRAS, 354,  675 	
\ref Contini, M., Viegas, S.M. 2001a, ApJS, 132, 211
\ref Contini, M., Viegas, S.M. 2001b, ApJS, 137, 75 
\ref Contini, M., Viegas, S. M., Prieto, M. A. 2002 A\&A,  386, 399 
\ref Contini, M., Radovich, M., Rafanelli, P., Richter, G.M. 2002, ApJ, 572, 124
\ref Dudik, R.P., Satyapal, S., Marcu, D. 2009, ApJ, 691, 1501
\ref Eisenhauer, G. et al. 2005, ApJ, 628, 246
%\ref Erickson, E. F., Colgan, S. W. J., Simpson, J. P., Rubin, R. H., Morris, M, Haas, M. R.
1991, ApJ, 370, L69
%\ref Figer, D. F. et al. 2002, ApJ, 581, 258
\ref Ghez, A.M. et al. 2005, ApJ, 620, 744
\ref Girimaji S.~S., Zhou Y., 1995, PhLA, 202, 279
\ref Goldman, I. 2000, ApJ, 54 
\ref Goldman, I.  2007, IAU Symposium 237, 96-1001, 701,arXiv:astro-ph/0703793
\ref Graham, M. J., Zhang, Q. 2000 ApJS, 127, 339
\ref Ho, L.C., Filippenko, A.V., Sargent, W.L.W. 1997 ApJS, 112, 315 
\ref Kaufman, M. J., Wolfire, M. G., Hollenbach, D. J., Luhman, M. L.  1999 ApJ,  527, 795  
\ref Kraemer, S. B.; Crenshaw, D. M.; Hutchings, J. B.; Gull, T. R.; Kaiser, M. E.; Nelson, C. H.; Weistrop, D.  2000, ApJ, 531, 278
\ref Kuraszkiewicz, J. K., Wilkes, B. J., Czerny, B., Mathur, S., Brandt, W. N.,
 Vestergaard, M. 2000 NewAR, 44, 573 
\ref Laursen, P.; Sommer-Larsen, J.; Andersen, A. C.  2009, ApJ, 704,164
%\ref Lesiurs, M. 1997, Turbulence in Fluids, \S 6.10
\ref Maiolino, R. et al. 2005, A\&A, 440, L51
\ref Maiolino, R., Caselli, P., De Zotti, G. 2009 SPICA Workshop.04004. EDP Sciences, 2009
\ref Medvedev M.~V., Silva L.~O., Kamionkowski M., 2007, AIPC, 932, 117 
\ref Miville-Desch{\^e}nes M.-A., Lagache G., Boulanger F., Puget J.-L., 2007, A\&A, 469, 595 
\ref Miville-Desch{\^e}nes M.~-., et al., 2010, arXiv, arXiv:1005.2746 
\ref Mikaelian K.~O., 1990, PhFl, 2, 592 
\ref Mininni P.~D., 2007, PhRvE, 76, 026316 
\ref Molinari,S. et al. 2010, A\&A, in press, arXiv:1005.3317
%\ref Morris, M., Yusef-Zadeh, F.  1987, ApJ, 320, 657
\ref Nagao, T., Marconi, A., Maiolino, R. 2006 A\&A, 447,157
\ref Nakanishi, K. \& Sofue, Y. 2006, PASJ, 58, 847
\ref  Passot T., Pouquet A., Woodward P., 1988, A\&A, 197, 228
\ref  Radovich, M., Rafanelli, P., Birkle, K.,  Richter, G. 1997, Astron. Nachr., 318, 229  
\ref Rubin, D. et al 2009, A\&A, 494, 647
\ref Sch\"{o}del, R. Bower, G,C., Muno, M.P., Nayakshin, S., Ott, T.
2006, Journal of Physics: Conference Series, Volume 54, Proceedings of "The Universe Under the Microscope - Astrophysics at High Angular Resolution",held 21-25 April 2008, in Bad Honnef, Germany. Editors: Rainer Schoedel, Andreas Eckart, Susanne Pfalzner and Eduardo Ros, pp. (2006).
\ref Simpson, J P.; Colgan, S. W. J., Cotera, A. S., Erickson, E. F.,
Hollenbach, D. J., Kaufman, M. J., Rubin, R. H. 2007, ApJ, 670, 1115
\ref Shields, J. C., \& Filippenko, A. V. 1990, AJ, 100, 1034 
\ref Schultheis, M., Sellgren, K., Ramírez, S., Stolovy, S., Ganesh, S., Glass, I. S.,  Girardi, L. 
2009, A\&A, 495, 157 
\ref  Shu, F.~H.\ 1992.\ Physics of Astrophysics, Vol. II. University Science Books.
\ref Spinoglio, L.  Malkan, M.A., Smith, H.A., Gonzalez-Alfonso, E., Fischer, J. 200, ApJ, 623, 123 
\ref  Stanimirovic S., Staveley-Smith L., Dickey J.~M., Sault R.~J., Snowden S.~L., 1999, MNRAS, 302, 417
\ref Vall{\'e}e J.~P., 2004, NewAR, 48, 763
\ref Veilleux, S., Osterbrock, D.E. 1987 ApJ,   63, 295
\ref Veilleux, S.; Kim, D.-C.; Sanders, D. B., Mazzarella, J. M., Soifer, B. T.  1995 ApJS,  98, 171 
\ref Walter, F. et al. 2009, Nature, 457, 699 
\ref Yusef-Zadeh, F., Morris, M.  1987, ApJ, 320, 557

\end{document}